\documentclass[journal=jpcafh,manuscript=article]{achemso}
\usepackage{longtable}
\usepackage{multirow}
\usepackage{amsmath}
\usepackage{subfigure}
\usepackage[usenames,dvipsnames]{xcolor}
\usepackage{soul}
\usepackage{bm}

\usepackage{amssymb}
\usepackage{siunitx}
\usepackage{threeparttable}
\usepackage{multicol} \usepackage{wrapfig} \usepackage{enumitem}
\usepackage{bm} \usepackage{outlines} 

\SectionNumbersOn

\author{Silvan K\"aser} \affiliation[University of Basel]{Department
  of Chemistry, University of Basel, Klingelbergstrasse 80 , CH-4056
  Basel, Switzerland}

\author{Jeremy O. Richardson} \affiliation[ETH Zurich]{Laboratory of
  Physical Chemistry, ETH Zurich, 8093 Zurich, Switzerland}

\author{Markus Meuwly} \affiliation[University of Basel]{Department of
  Chemistry, University of Basel, Klingelbergstrasse 80 , CH-4056
  Basel, Switzerland} \email{m.meuwly@unibas.ch}

\title{Transfer learning for affordable and high quality tunneling
  splittings from instanton calculations}
\begin{document}
\date{\today}

\begin{abstract}
The combination of transfer learning (TL) a low level potential energy
surface (PES) to a higher level of electronic structure theory
together with ring-polymer instanton (RPI) theory is explored and
applied to malonaldehyde.  The RPI approach provides a semiclassical
approximation of the tunneling splitting and depends sensitively on
the accuracy of the PES. With second order M{\o}ller-Plesset
perturbation theory (MP2) as the low-level (LL) model and energies and
forces from coupled cluster singles, doubles and perturbative triples
(CCSD(T)) as the high-level (HL) model, it is demonstrated that
CCSD(T) information from only 25 to 50 judiciously selected structures
along and around the instanton path suffice to reach HL-accuracy for
the tunneling splitting. In addition, the global quality of the HL-PES
is demonstrated through a mean average error of 0.3 kcal/mol for
energies up to 40 kcal/mol above the minimum energy structure (a
factor of 2 higher than the energies employed during TL) and $< 2 $
cm$^{-1}$ for harmonic frequencies compared with computationally
challenging normal mode calculations at the CCSD(T) level.
\end{abstract}

\section{Introduction}
\begin{figure}[h!]
\centering
\includegraphics[width=0.75\textwidth]{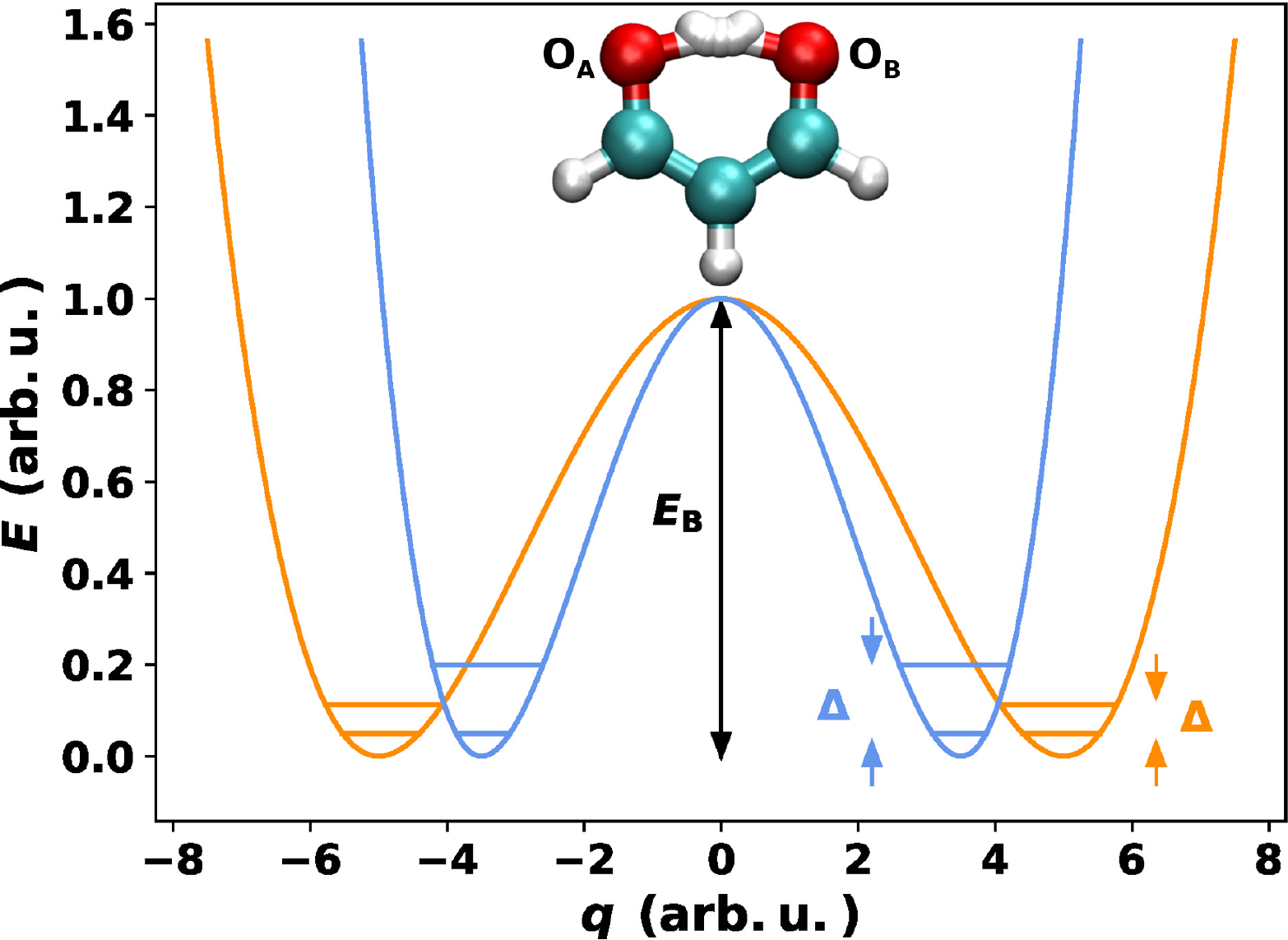}
\caption{Schematic illustration of two symmetric double-well
  potentials with the same energy barrier $E_{\rm B}$ but differing
  curvatures and tunneling splittings $\Delta$. Malonaldehyde and its
  instanton path are shown.}
\label{fig:schematic}
\end{figure}

Tunneling splittings are exquisitely sensitive to the accuracy of a
molecular potential energy surface (PES). The nuclear wave-functions
corresponding to the two or multiple quantum mechanical bound states
involved in the split energy levels probe an extended region on the
underlying PES. Furthermore, the tunneling splitting also informs
about the barrier height and the shape of the PES in the region
connecting the two wells, see Figure~\ref{fig:schematic}. Due to all
the above, tunneling splittings constitute a meaningful and stringent
test of the level of theory at which the underlying PESs were
calculated and the accuracy of their representation required for
simulations from which the splittings are determined.\\

\noindent
Even if a PES is given, accurate computation of tunneling splittings
for multidimensional systems from quantum-based methods itself is a
formidable task. The ring-polymer instanton (RPI) approach provides a
semiclassical approximation of a tunneling process and can be used to
calculate tunneling splittings in molecular
systems.\cite{tunnel,InstReview,hexamerprism,richardson2017full} As
was shown for the formic acid dimer\cite{richardson2017full}, it is
necessary to include all degrees of freedom of the molecule for a
quantitative comparison with experiment. This often means that the
(semiclassical) full-dimensional instanton approximation is more
accurate than a reduced-dimensional quantum calculation. Instanton
theory is based on the path-integral formulation of quantum mechanics
and is formally exact only in the limit of $\hbar \rightarrow
0$.\cite{Uses_of_Instantons} In many previous studies it has been
shown to give predictions within about 20\% of fully
quantum-mechanical calculations using the same PES for typical
molecular systems, as long as the barrier height is significantly
higher than the zero-point energy along the tunneling
mode.\cite{tunnel,richardson2017full,jahr2020instanton} Instanton
calculations which, contrary to exact quantum calculations such as
wave-function
propagation,\cite{Hammer2011malonaldehyde,Schroeder2011malonaldehyde}
scale well with system size, are often used in combination with
analytical, full-dimensional PESs. Path-integral molecular dynamics
(PIMD) simulations also scale well with system size but are
considerably more expensive than an instanton
calculation.\cite{Matyus2016tunnel1,Vaillant2018dimer}\\

\noindent
In principle it is possible to implement the instanton approach using
\textit{on-the-fly} \textit{ab-initio} electronic structure
calculations.\cite{Milnikov2003,chiral} However, because energies,
gradients and Hessians are needed for each ring-polymer bead, this may
be impractical for medium-sized molecules if high accuracy from
coupled cluster with perturbative triples (CCSD(T)) level of theory is
sought. More recent work has been devoted to combining machine
learning (ML) and instanton theory to reduce overall computational
expense. Gaussian process regression (GPR) has been used to obtain a
local fit of the PES around the dominant tunneling pathway to
calculate rate constants.\cite{laude2018ab} It has been shown that the
GPR rate constants are on par with the \textit{ab initio} results,
however, reducing the number of required electronic structure
calculations by an order of magnitude. Similarly, instanton rate
theory has been combined with NNs to obtain the PES more efficiently
as compared to the \textit{on-the-fly}
approach\cite{cooper2018potential,McConnell2019instanton}. \\

\noindent
As an alternative, full-dimensional PESs can now be constructed for
medium-sized molecules from which tunneling splittings can also be
determined using the instanton approach. The generation of ML PESs
based on large data sets of \textit{ab initio} data is a challenging
task\cite{unke2020high} and accuracy as well as the level of theory of
the PES is of crucial importance for the accurate determination of
tunneling splittings. The ``gold standard'' CCSD(T) approach scales as
$N^7$ (with $N$ being the number of basis
functions)\cite{friesner2005ab} which quickly becomes computationally
prohibitive for generating data to build a full-dimensional PESs even
for relatively small molecules. To avoid the need for calculating
large \textit{ab initio} data sets at high levels of theory transfer
learning (TL)
\cite{pan2009survey,taylor2009transfer,smith2019approaching} and
related $\Delta$-ML\cite{DeltaPaper2015} were shown to be data and
cost-effective
alternatives\cite{mm.ht:2020,mm.vibratingh2co:2020,nandi2021delta,mm.anharmonic:2021,qu2021breaking,kaser2022transfer}.\\

\noindent
The combination of TL and instanton theory appears particularly
appealing as the instanton path (IP) can be determined on a low-level
PES, which gives a rough approximation to the true tunneling path, and
can be included (and iteratively refined if needed) into the TL data
set. Additionally, the IP is inherently local and, thus, allows
concentrating on improving only a small part of a PES. While instanton
theory has been used in combination with ML
schemes\cite{laude2018ab,Muonium,McConnell2019instanton}, the present
work demonstrates the first combination of instanton theory with
TL. The capability of the combined approach is demonstrated for the
extensively studied malonaldehyde system exhibiting intramolecular
hydrogen transfer (HT).\\

\noindent
Ring Polymer Instanton theory has been employed to calculate the
tunneling splitting of malonaldehyde on a permutationally invariant
polynomial (PIP) PES fitted to 11147 near basis-set-limit frozen-core
CCSD(T) electronic energies\cite{wang2008full}. The splitting was
found to be 25~cm$^{-1}$ with RPI \cite{jahr2020instanton} as well as
with a strongly related instanton method
\cite{cvitas2016locating}. The same PES was also used to calculate
tunneling splittings using the \textit{fixed-node} diffusion Monte
Carlo (DMC) method giving 21.6~cm$^{-1}$ with a statistical
uncertainty of 2 to 3~cm$^{-1}$.\cite{wang2008full} For DT, computed
values on the PIP PES are 3.3 and 3.4~cm$^{-1}$ using
RPI\cite{jahr2020instanton} and the related instanton
method\cite{cvitas2016locating}, and $3.0\pm2-3$~cm$^{-1}$ from DMC
simulations.\cite{wang2008full} The tunneling splitting from RPI
calculations on a LASSO fit to CCSD(T)(F12*) energies was found to be
19.3~cm$^{-1}$.\cite{jahr2020instanton,mizukami2014compact} To
validate such computations, direct comparison with experiment is also
of interest. Reliable tunneling splittings from experiment are only
available for a few select
systems\cite{firth1991tunable,baba1999detection,redington2008infrared,murdock2010vibrational,ortlieb2007proton,duan:2017,caminati:2019,insausti2022rotational}.
For malonaldehyde, the experimentally determined tunneling splitting
is 21.583~cm$^{-1}$ and 2.915~cm$^{-1}$ for HT and deuterium transfer
(DT),
respectively\cite{firth1991tunable,baba1999detection,baughcum1984microwave}. Although
these results are not in perfect agreement with experiment, they are
close enough that spectroscopic assignments can be made and provide
detailed mechanistic information about the tunneling process. \\

\noindent
The purpose of the present work is to develop and quantitatively
assess an evidence-based procedure to determine reliable tunneling
splittings by combining transfer learned PESs with instanton
calculations. It is shown that this can dramatically reduce the cost
of the overall simulation in comparison to working with \emph{ab
  initio} potentials. The work is structured as follows. First, the
methods and the generation of the data sets are presented. This is
followed by a thorough evaluation of the accuracy of the transfer
leaned PESs in terms of tunneling splittings and harmonic
frequencies. Finally, the results are discussed and conclusions are
drawn.\\

\section{Methods}
\subsection{Ring-Polymer Instanton Theory}
In a one-dimensional model, instanton theory is strongly related to
the WKB approximation\cite{Milnikov2001}.  Its main advantage,
however, is that it can also be applied to multidimensional systems,
in which it locates the uniquely defined optimal tunneling
pathway.\cite{Perspective} This pathway, known as the instanton, is
defined as a long imaginary-time $\tau\rightarrow\infty$ path
connecting two degenerate wells which minimizes the action, $S$.  In
computations, the path is located using an efficient ring-polymer
optimization based on discretizing the path into $N$ ring-polymer
beads and taking the limit $N\rightarrow\infty$ (typically on the
order of 1000 is sufficient for convergence).  The action is
determined by the distance between neighbouring beads as well as the
potential-energy of each bead, i.e. it uses information \emph{along}
the IP\@.  Full technical details are presented in previous
work.\cite{tunnel,InstReview} In general, the IP is not equivalent to
the minimum-energy pathway (MEP) and will not even pass through a
saddle point.  This is because the instanton finds a compromise
between length and height to optimize the tunneling path.  Unlike PIMD
or DMC, no random numbers or statistical errors are involved and so
the instanton (once it is converged with $\tau\rightarrow\infty$ and
$N/\tau\rightarrow\infty$) is in principle uniquely determined by the
PES.\\

\noindent
Once the IP has been located, fluctuations around the path are
computed to second order and the information is combined into the term
$\Phi$, i.e. this is based on information \emph{around} the IP\@.  For
this, one requires the Hessians (second-derivative matrix of PES) at
each bead. The final prediction for the tunneling splitting (in a
double-well system) is given by
\begin{align}
    \Delta = \frac{2\hbar}{\Phi} \sqrt\frac{S}{2\pi\hbar} \,
    \mathrm{e}^{-S/\hbar}.
\label{eq:rpi_splitting} 
\end{align}
Because $S$ appears in the exponent it is particularly important to
determine this quantity with high accuracy.

\noindent
The method has also been generalized to treat tunneling in systems
with multiple (more than two) wells \cite{water,hexamerprism} and in
cases with non-degenerate wells for instance due to asymmetric
isotopic substitution.\cite{jahr2020instanton} The approach outlined
in this work is, in principle, also applicable to these extensions.\\

\subsection{Machine Learning}
All PESs used in this work are represented with a high-dimensional
neural network (NN) of the PhysNet\cite{MM.physnet:2019}
architecture. PhysNet is a `message-passing'\cite{gilmer2017neural} NN
that employs learnable descriptors of the atomic environments to
predict individual atomic energy contributions $E_i$ and partial
charges $q_i$. The descriptors are initialized as ${\bf x}_i^0 = {\bf
  e}_{Z_i}$, where ${\bf e}_{Z_i}$ corresponds to a parameter vector
defined by the nuclear charge $Z_i$, i.e. atoms of the same element
share the same descriptor. The descriptor is then iteratively updated
and refined to best describe the local chemical environment of each
atom $i$ by passing `messages' between atoms within a cut-off $r_{\rm
  cut}$ following
\begin{align}
    {\bm x}_i^{l+1} = {\bm x}_i^l + \sum_{r_{ij} < r_{cut}}
    \mathcal{F} ({\bm x}_i^l, {\bm x}_j^l, r_{ij}).
\end{align}
where $r_{\rm cut}$ was 10~\AA. Here, ${\bm x}_i^l$ and ${\bm x}_j^l$
are the descriptors of atoms $i$ and $j$ at iteration $l$, $r_{ij}$ is
their interatomic distance and $\mathcal{F} ({\bm x}_i^l, {\bm x}_j^l,
r_{ij})$ is the `message-passing' function (for details see
Ref.~\citenum{MM.physnet:2019}). Because only pairwise distances are
used to encode the atoms' chemical environment and summation is
commutative, the resulting descriptors (and thus the PES) are
invariant with respect to translation, rotation and permutation of
identical atoms, which is of particular importance when describing
tunneling between degenerate wells. The descriptors are then used to
predict partial charges $q_i$ (which are corrected to ensure total
charge conservation) and the total energy of the chemical system by
summation of the atomic contributions and explicitly including
long-range electrostatics according to
\begin{align}\label{eq:energy_expression}
    E = \sum_i E_i + k_e\sum_{i=1}^N\sum_{j>i}^N \frac{q_iq_j}{r_{ij}}
\end{align}
Here, $k_e$ represents Coulomb's constant and the second term
involving $\frac{q_iq_j}{r_{ij}}$ is damped to avoid numerical
instabilities caused by the singularity at $r_{ij} = 0$ (for details
refer to Ref.~\citenum{MM.physnet:2019}). The forces ${\bm F}$ and
Hessians ${\bm H}$ can be obtained analytically using reverse mode
automatic differentiation\cite{baydin2017automatic} as implemented in
Tensorflow\cite{abadi2016tensorflow}.\\

\noindent
The learnable parameters of PhysNet are fitted to reference \textit{ab
  initio} energies, forces and dipole moments following the strategy
outlined in Reference~\citenum{MM.physnet:2019}. The partial charges
$q_i$ are fitted to the \textit{ab initio} dipole moment ($\mu =
\sum_i^Nq_i{\bm r_i}$) and explicitly enter the energy expression (see
equation~\ref{eq:energy_expression}). In the present work, the TL
scheme is employed whereby the parameters of a low-level (LL)
treatment are used as a meaningful initial guess and are fine-tuned
using higher-level information. For TL, the learning rate is reduced
from $10^{-3}$ (as for learning a model from scratch) to
$10^{-4}$. The LL in the present work is the full-dimensional PES for
malonaldehyde at the MP2/aug-cc-pVTZ level of theory (henceforth,
PhysNet MP2 PES) which is available from previous work and was trained
on $\sim 70000$ reference structures.\cite{mm.ht:2020} This PhysNet
PES has a barrier for HT of 2.79~kcal/mol which compares to a
reference value of 2.74~kcal/mol calculated at the MP2/aug-cc-pVTZ
level of theory and the reference harmonic frequencies are reproduced
with a root-mean-square deviation (RMSD) of 3.6~cm$^{-1}$. The
high-level (HL) treatment is the considerably higher and
computationally much more demanding CCSD(T)/aug-cc-pVTZ level of
theory at which energies, forces and dipole moments are calculated
using Molpro\cite{MOLPRO} for all data points used in TL.\\

\subsection{Data Set Generation}
Transfer learning requires high-level energies, forces and dipole
moments for selected geometries of the system considered and ideally
cover all spatial regions relevant for the observable(s) of interest.
Without additional {\it a priori} information it is advantageous to
generate an initial pool of structures which can be used for TL to
fine-tune the LL treatment. When selecting configurations, it is not
necessary to sample both potential wells since PhysNet handles this
symmetry by construction. Here, the initial pool contained 862
malonaldehyde configurations consisting of:
\begin{itemize}
    \item 111 geometries along the MEP of the PhysNet MP2 PES.
    \item 110 geometries along the IP of the PhysNet MP2 PES.
    \item 111 geometries along the IP determined on a PES that was
      transfer learned by using CCSD(T) information of the 111 MEP
      geometries (see above) to have an energy barrier closer to the
      \textit{ab initio} CCSD(T) barrier.
    \item 280 geometries obtained from normal mode sampling (NMS)
      around the equilibrium geometry. For this purpose, normal mode
      vectors and corresponding force constants are determined
      \textit{ab initio} at the MP2/aug-cc-pVTZ level of theory.
    \item 240 geometries around the IP as obtained from NMS.
    \item 10 geometries along the IP of TL$_1$ (see
      Section~\ref{sec:smalltls}).
\end{itemize}

\noindent
This data set is referred to as the ``Extended Data Set'' and transfer
learned models using it are called TL$_{\rm ext}$. To probe the
dependence of barrier heights and tunneling splittings on details of
the training, ten independent models were trained on different splits
of the data for TL$_{\rm ext}$ (and all the subsequent TLs). For each
of the ten resulting PESs an instanton calculation was carried
out. From this information, averages and standard deviations for the
barrier heights and tunneling splittings were determined.\\

\noindent
After validating the performance of TL$_{\rm ext}$ from instanton
calculations on each of the independently trained models, smaller
subsets of the Extended Data Set were selected, employed for TL and
subsequent tunneling splitting calculations.\\

\section{Results}
To set the stage, the tunneling splittings for malonaldehyde were
calculated on the PhysNet MP2 PES using RPI theory. The tunneling
splitting calculations were carried out with three different values of
the imaginary time, $\tau$, corresponding to effective `temperatures'
$T=\hbar/k_B\tau \in [50, 25, 12.5]$~K and with different numbers of
beads $N \in [2^5,..,2^{12}]$ to ensure convergence.  Formally the
instanton result is defined in the low-temperature limit, which is
equivalent to infinitely-long imaginary times. The results are
summarized in Table~S1. A tunneling splitting
of 96~cm$^{-1}$ is obtained compared with 25 cm$^{-1}$ from instanton
calculations\cite{cvitas2016locating,jahr2020instanton} on the
PIP-representation\cite{wang2008full} of the CCSD(T) reference data
and 21.6 cm$^{-1}$ from
experiments.\cite{firth1991tunable,baba1999detection,baughcum1984microwave}
This illustrates the insufficient quality of the MP2 level of theory
to capture tunneling splittings correctly. For the following, all
instanton calculations were carried out with $N = 4096$ beads at an
effective temperature of $T = 25$~K, which was found to be more than
sufficient for convergence of $\Delta$ to two significant figures.\\

\begin{figure}[h!]
\centering \includegraphics[width=1\textwidth]{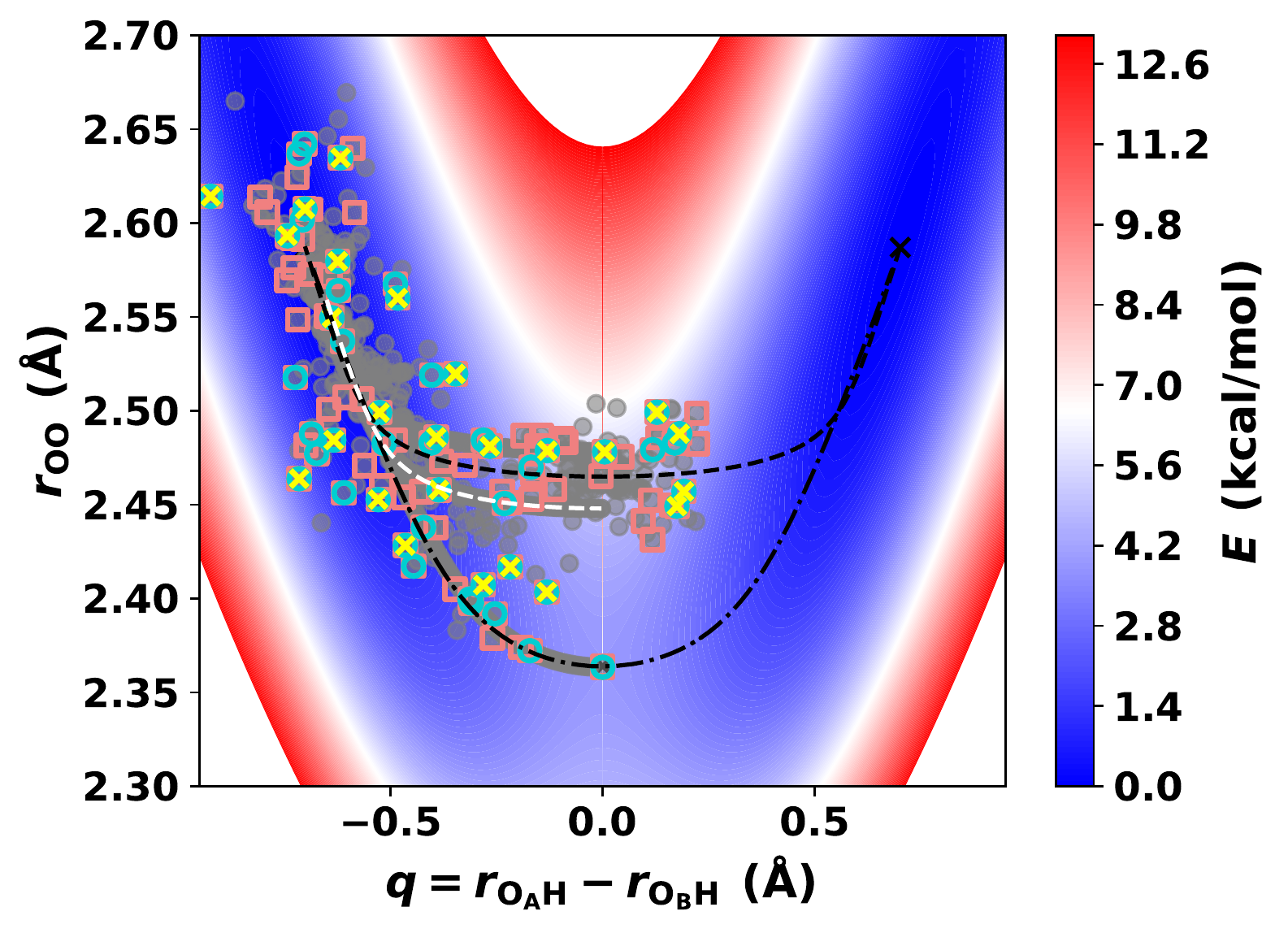}
\caption{Data sets used for TL projected onto a 2D cut through the
  TL$_{\rm ext}$ PES spanned by the O--O distance and the reaction
  coordinate $q = r_{\rm O_AH} - r_{\rm O_BH}$, for labels see
  Fig.~\ref{fig:schematic}. The Extended Data Set (862 structures,
  gray circles) is shown together with sets for TL$_0$ (25 structures,
  yellow crosses), TL$_1$ (50 structures, turquoise circles), and
  TL$_2$ (100 structures, salmon squares). The MEP and the instanton
  path (as determined on the TL$_{\rm ext}$ PES) are marked with a
  dash-dotted and a dashed line, respectively. The IP on the PhysNet
  MP2 PES is the white dashed line and clearly differs from that
  (black dashed) on the CCSD(T) PES.}
\label{fig:datasetplot}
\end{figure}

\subsection{Performance of TL$_{\rm ext}$}
As a reference for the following exploration, the performance of
TL$_{\rm ext}$ using the full set of 862 energies, forces and dipole
moments determined at the CCSD(T)/aug-cc-pVTZ level is first
assessed. These geometries are shown as a projection onto the PES
spanned by the O--O distance and the reaction coordinate $q = r_{\rm
  O_AH} - r_{\rm O_BH}$ in Figure~\ref{fig:datasetplot} (gray
circles). As the PES is symmetric with respect to $q = 0$, the same
geometry appears to the left and to the right of the mirror plane.
The Extended Data Set was split according to $\sim 80/10/10$~\% into
training/validation/test sets, from which the test sets were used only
for testing. Across the 10 TL models, the separate test sets were
predicted on average with MAE($E$) $\approx 0.006$, RMSE($E$) $\approx
0.009$~kcal/mol, MAE($F$) $\approx 0.03$ and RMSE($F$) $\approx
0.07$~kcal/mol/\AA. The average barrier height on the ten
transfer-learned PESs ($\langle{\rm TL}_{\rm ext}\rangle$) was $E_{\rm
  B} = 3.8945 \pm 0.0006$ kcal/mol which compares with an \textit{ab
  initio} barrier of 3.8948~kcal/mol determined at the
CCSD(T)/aug-cc-pVTZ level of theory determined from present
calculations. The RPI tunneling splittings for HT and DT were
$\Delta_{\rm H} = 25.3 \pm 0.2$ cm$^{-1}$ and $\Delta_{\rm D} = 3.7
\pm 0.03$ cm$^{-1}$, respectively, see Table
\ref{tab:all_tl_splittings} and S2. These
results compare with computed splittings on the PIP representation of
CCSD(T)/aug-cc-pVTZ reference calculations using instanton
calculations that yield 25/3.4
cm$^{-1}$\cite{cvitas2016locating,jahr2020instanton} and experimental
splittings of
21.6/2.9~cm$^{-1}$.\cite{firth1991tunable,baba1999detection,baughcum1984microwave}
Hence, the transfer-learned PES using fewer than 1000 higher level
CCSD(T)/aug-cc-pVTZ energies and forces together with the same method
for determining the tunneling splitting performs on par with
calculations on the PIP representation of the $\sim 11000$
CCSD(T)/aug-cc-pVTZ energies.\cite{wang2008full} Based on this it is
of much practical interest to further reduce the number of HL
calculations required to achieve the same result. Therefore, in a next
step, different subsets of the Extended Data Set are considered and
used for TL to arrive at an ideally small number of HL points while
still retaining the accuracy in tunneling splittings from instanton
calculations.\\

\begin{table}[h!]
\begin{tabular}{c|c|c|ccc}
& $N_{\rm data}$ & \textbf{$E_{\rm B}$} & \textbf{$\Delta_{\rm H}$} &\textbf{$S/\hbar$} &\textbf{$\Phi$}\\\hline
MP2 & 70k & 2.7889 & 96.3 & 4.502 & 42.770\\\hline
$\langle{\rm TL}_0\rangle$& 25&	$3.8925\pm	0.0119$&	$23.4\pm1.9$&	$5.749\pm0.026$	&	$57.405\pm4.486$\\
$\langle{\rm TL}_1\rangle$	& 50&	$3.8974\pm0.0161$	&	$24.9\pm1.1$	&	$5.764\pm0.012$	&	$53.096\pm2.375$\\
$\langle{\rm TL}_2\rangle$& 100&	$3.8941\pm0.0021$	&	$25.2\pm0.5$	&	$5.748\pm0.005$	&	$53.053\pm1.156$\\
$\langle{\rm TL}_{\rm ext}\rangle$& 862	&	$3.8945\pm0.0006$	&	$25.3\pm0.2$	&	$5.743\pm0.002$	&	$53.241\pm0.454$\\
\end{tabular}
\caption{Energy barriers $E_\mathrm{B}$ (kcal/mol), tunneling
  splittings $\Delta$ (in cm$^{-1}$) at $T = 25$~K and with $N =
  4096$, action $S/\hbar$ and fluctuation factor $\Phi$ (in a.u. of
  time $\hbar / E_h$) for malonaldehyde determined from TL$_{\rm ext}$ and
  TL$_{0,1,2}$ PESs. The {\it ab initio} barrier height for HT at the CCSD(T)/aug-cc-pVTZ level of theory   is 3.8948~kcal/mol.}
\label{tab:all_tl_splittings}
\end{table}

\subsection{Performance on Smaller Datasets: TL$_0$, TL$_1$, and TL$_2$}\label{sec:smalltls}
From the Extended Data Set containing 862 geometries, different
subsets were extracted. The size of the data set has to be chosen
small enough for efficient computation but sufficiently large to still
cover the appropriate regions of configurational space probed by the
instanton calculation.\\

\noindent
{\it TL$_0$:} To check whether a considerably smaller data set
suffices as a starting point, ten TLs were preformed on a data set
containing only 25 geometries: a) 5 IP geometries (approximately
equally spaced) determined on a PES that was transfer learned to have
a barrier closer to the \textit{ab initio} CCSD(T) barrier; b) 10
geometries, each, selected from the NMS around the equilibrium
geometry and the IP. This can be done by selecting geometries based on
a RMSD criterion on the $n$ atomic positions
($\sqrt{\frac{1}{n} \sum_{i=1}^n||v_i - w_i||^2}$ where $v_i$ and $w_i$
are two sets of Cartesian coordinates of atom $n$),  which was done as follows for both groups of geometries
that were generated with NMS. Starting from a random geometry, new
geometries are added iteratively if the RMSD with respect to the
selected ones is larger than a threshold. For this reason, the
threshold is maximized to include 10 geometries. Note that no MEP
geometries are added. The data set for TL$_0$ are the yellow crosses
in Figure~\ref{fig:datasetplot}.\\

\noindent
With this smallest subset the barrier height of the (ensemble of the)
transfer learned PES $\langle{\rm TL}_0\rangle$ is $E_{\rm B} = 3.8925
\pm 0.012$~kcal/mol which is, within errors, close to the target value
of 3.8948 kcal/mol determined at the CCSD(T)/aug-cc-pVTZ level. From
10 independent instanton calculations the average splitting is
$\Delta_{\rm H} = 23.4 \pm 1.9$ cm$^{-1}$ which is $\sim 2$ cm$^{-1}$
below that from the simulations on the TL$_{\rm ext}$ PES but still
within statistical fluctuation. Comparing the action $S/\hbar$ of the
IPs on TL$_0$ and TL$_{\rm ext}$ shows that they are comparable
($5.743 \pm 0.002$ vs. $5.749 \pm 0.026$) and even identical within
the error bars. However, the uncertainty on TL$_0$ is larger by an
order of magnitude compared with that on TL$_{\rm ext}$. Thus, the
action, $S/\hbar$, of the path is clearly less well defined on
TL$_0$. For the fluctuation factor $\Phi$ the differences are
considerably larger between the two families of PESs. Still, the
values themselves are within error bounds but again, the fluctuation
around the mean for TL$_0$ is ten times larger than that for the
TL$_{\rm ext}$ PESs. This conclusion also hold for DT, see
Table~S2. Overall, using only 25
additional data points as done for TL$_0$ already yields encouraging
results for the barrier height and tunneling splittings. To explore
further improvements new points were added and the process was
repeated.\\

\begin{figure}[h!]
\centering
\includegraphics[width=0.8\textwidth]{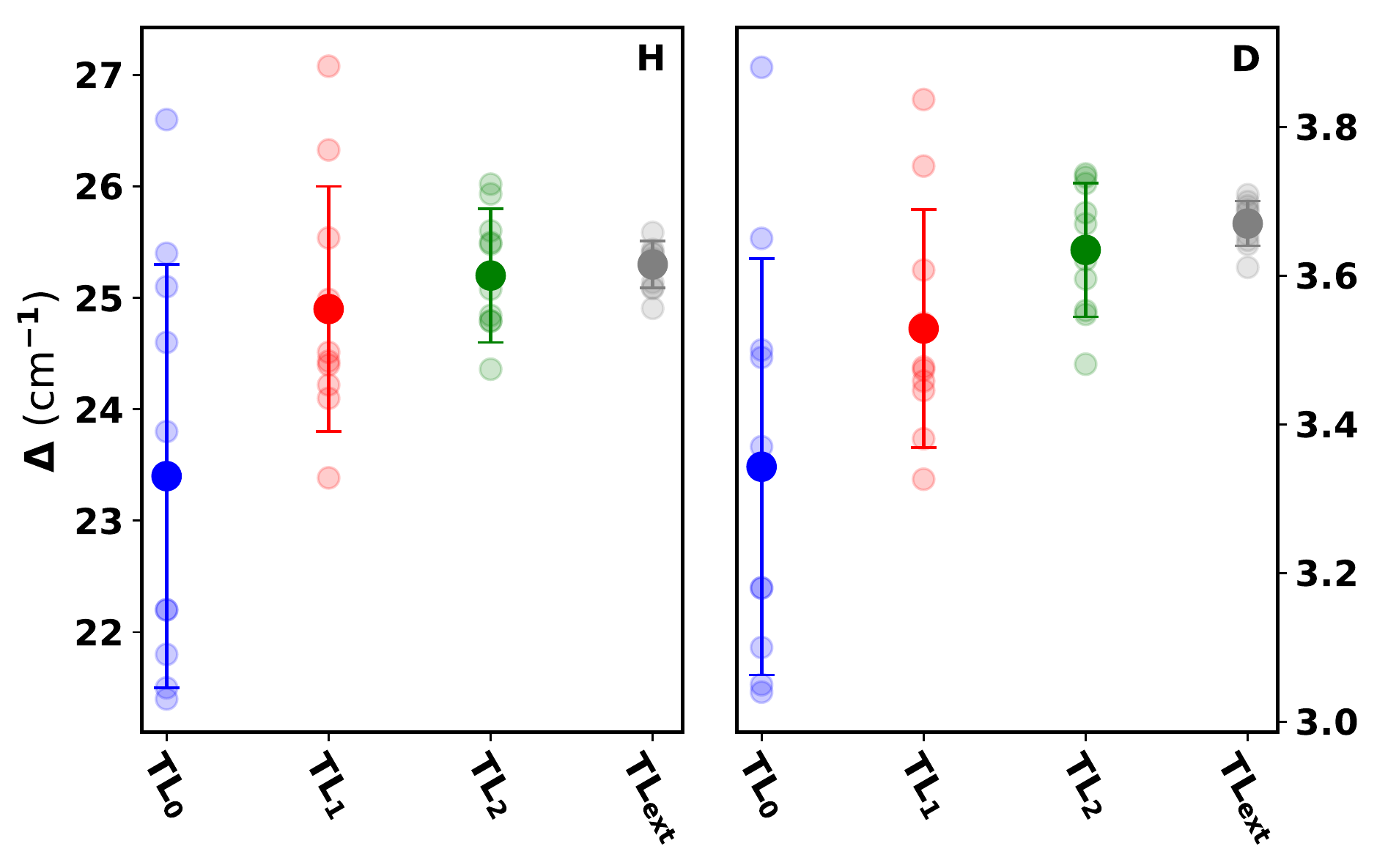}
\caption{Tunneling splittings for H- and D-transfer (left and right
  panels) from all TL PESs (transparent circles). The corresponding
  averages (opaque circle) and standard deviations (error bars as $\pm
  \sigma$) as obtained from TL$_0$ (blue), TL$_1$ (red), TL$_2$
  (green) and TL$_{\rm ext}$ are reported, too. Comprehensive
  lists of the tunneling splittings for all TL PESs are given in
  Tables~S3 to S6.}
\label{fig:splittings}
\end{figure}

\noindent
{\it TL$_1$:} For TL$_1$ the points used for TL$_0$ were extended and
increased to 50 points, see turquoise circles in
Figure~\ref{fig:datasetplot}. The data set for TL$_1$ contained: a) 5
geometries along the MEP of the PhysNet MP2 PES; b) 5 geometries along
the IP of a transfer learned PES from using the MEP points calculated
at CCSD(T); c) 20 geometries, each, selected from the NMS around the
equilibrium geometry and the IP\@. The MEP and IP geometries are
chosen with a uniform spacing along the respective path and the
geometries from NMS are selected following the RMSD approach outlined
for TL$_0$.\\

\begin{figure}[h!]
\centering
\includegraphics[width=0.8\textwidth]{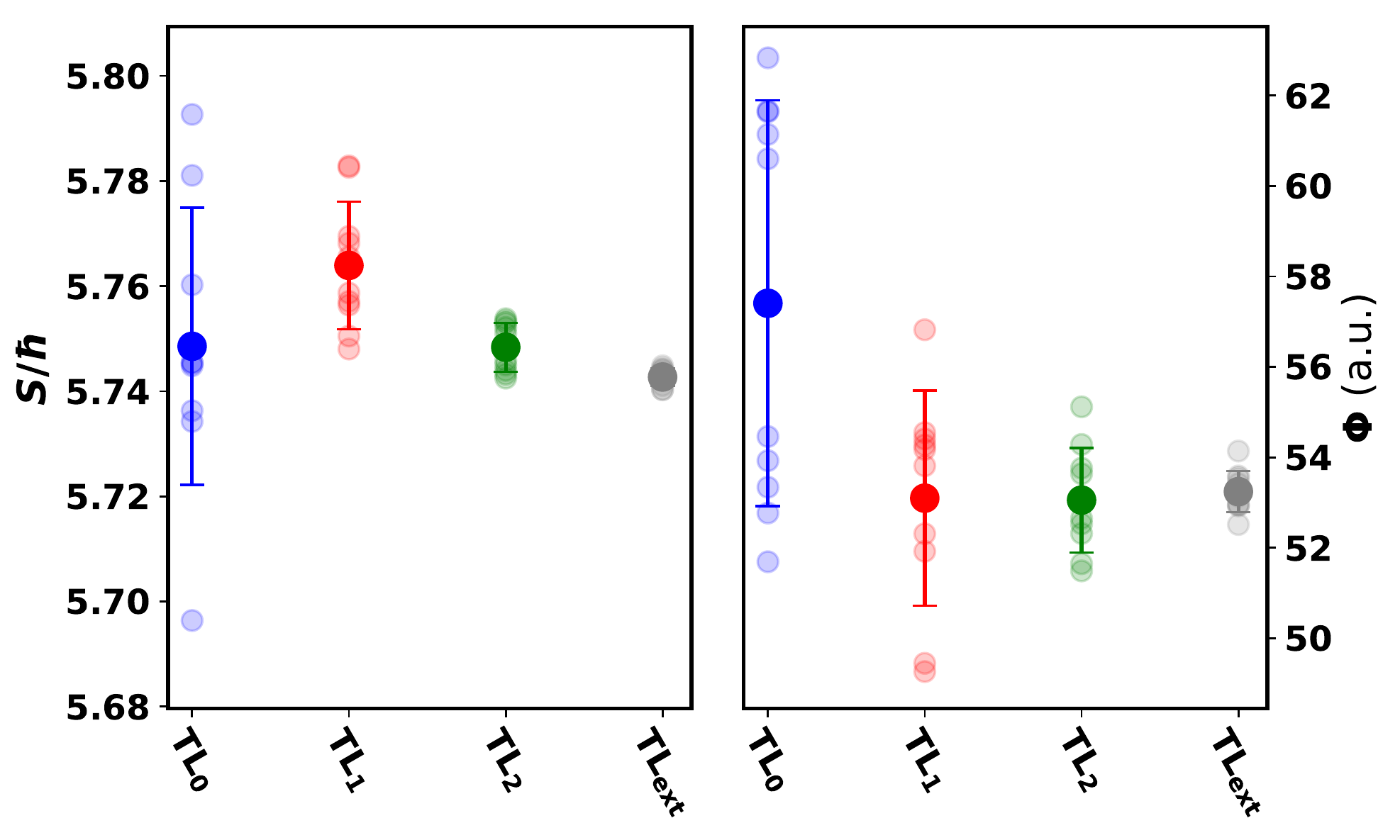}
\caption{Action $S/\hbar$ (left panel) and fluctuation factor $\Phi$
  (right panel) from all TL PESs (transparent circles) for HT
  (for DT see Figure~S5).
  The corresponding averages (opaque circle) and standard
  deviations (error bars as $\pm \sigma$) as obtained from TL$_0$
  (blue), TL$_1$ (red), TL$_2$ (green) and TL$_{\rm ext}$ are
  indicated as well.}
\label{fig:action}
\end{figure}

\noindent
TL$_1$ yields an averaged barrier height $E_{\rm B} = 3.8974 \pm
0.0161$ which agrees within error bars with that of TL$_{\rm ext}$ and
the \textit{ab initio} value (3.8948~kcal/mol). The splitting
$\Delta_{\rm H} = 24.9 \pm 1.1$ cm$^{-1}$ is only 0.4 cm$^{-1}$ below
that of TL$_{\rm ext}$. The key improvement is that the fluctuation
factor $\Phi$ agrees considerably better with TL$_{\rm ext}$ than for
TL$_0$ and the remaining 2\% discrepancy can be traced to the $0.02$
absolute difference in $S/\hbar$.  Overall, increasing the number of
geometries used for TL to $N_{\rm data} = 50$ in this fashion leads to
a PES which reproduces the barrier height and splittings from TL$_{\rm
  ext}$.  To further probe convergence of these results yet a larger
data set was considered.\\

\noindent
{\it TL$_2$:} While the accuracy of TL$_1$ might appear satisfactory,
convergence of TL$_1$ cannot be checked without TL$_{\rm ext}$. Thus,
TL$_1$ was further extended to yield TL$_2$. This was accomplished
from a strategy related to adaptive
sampling.\cite{behler2015constructing} Two independent models from
TL$_1$ were used to predict the energy $E$ of the remaining pool of
geometries obtained from NMS. From these geometries, the 40 geometries
with the largest deviation (here $\sim 0.02$~kcal/mol) between the
prediction of the two NNs are added to the data set. If a large
deviation between the energy predictions of the two models is found,
it is likely that no or too little reference data has been included in
TL$_1$. In addition, 5 geometries along the MEP and instanton of the
TL$_1$ were added such that TL$_2$ contained a total of 100 geometries
(see salmon squares in Figure~\ref{fig:datasetplot}).\\

\noindent
With TL$_2$ the barrier height $E_\mathrm{B}$, splitting $\Delta$,
action $S/\hbar$ and fluctuation $\Phi$ further improve over those
from TL$_1$ and are closer to the results from TL$_{\rm ext}$, see
Table \ref{tab:all_tl_splittings} and S2.
Within $1 \sigma$ all values for HT and DT agree with those from
TL$_{\rm ext}$ and with the reference from the literature except for
the tunneling splitting for DT. Overall, addition of 50 to 100
additional points from the HL treatment appears to suffice to arrive
at a quantitatively correct PES transfer-learned from the LL treatment
(MP2/aug-cc-pVTZ). Figure~\ref{fig:splittings} illustrates the gradual
convergence for TL$_0$ to TL$_2$ towards the results found for
TL$_{\rm ext}$, which is promising. While exceptional agreement
regarding the energy barrier $E_\mathrm{B}$ (as the ``simpler''
property) is found for all the TLs the standard deviation of the
splittings $\Delta_{\rm H}$ (which is more challenging to obtain) can
be reduced from 1.9~cm$^{-1}$ (7.5\%) for TL$_0$ to 0.5~cm$^{-1}$
(2\%) for TL$_2$. The results from TL$_2$ are accurate to within
0.1~cm$^{-1}$ for HT and 0.04~cm$^{-1}$ for DT as compared to ${\rm
  TL}_{\rm ext}$. This corresponds to deviations of 0.4\% and
1.1\%. The IPs themselves are reported in
Figures~S1 to S4 and show
slight deviations between the different TLs on the smallest data set
(TL$_0$), but for TL$_1$, TL$_2$ and TL$_{\rm ext}$ they are hardly
distinguishable.\\

\noindent
In summary, it has been found that with between 25 and 50 HL energies
and forces for judiciously chosen structures the correct barrier
height, tunneling splitting, action and fluctuation can be obtained,
see Table \ref{tab:all_tl_splittings}. This is the foundation to
further optimize the procedure, ideally based entirely on information
available from the LL surface by minimizing the amount of data
required from the HL treatment.\\

\subsection{Towards an Optimized Procedure}
For an even more efficient procedure, an approach is sought that is
based on information about the LL-PES only. Hence, an attempt is made
to further reduce the computational effort by minimizing the number of
structures for which HL calculations need to be carried out for an
improved PES and tunneling splittings. Therefore, it is explored which
elements of the procedure are most important for obtaining high
accuracy to cost ratios. For moving towards a more evidence-guided,
optimized procedure, TL is carried out from LL-information that is
only contained in the MEP and IP as follows: a) only the MEP (TL$_{\rm MEP}$) using
111 geometries along the MEP of the PhysNet MP2 PES; b) only the IP
(TL$_{\rm IP}$) using 110 geometries along the IP of the PhysNet MP2
PES; and c) a combination of the MEP and the IP (TL$_{\rm MI}$)
including 111 MEP and 110 instanton geometries as obtained from the
PhysNet MP2 PES. The HL information for TL consisted again of
energies, forces and dipole moment determined at the
CCSD(T)/aug-cc-pVTZ level of theory. A total of 5 TLs were performed,
each on different splits of the data. The two best NNs (judged from
the performance on the validation set) were used for further
analysis.\\

\begin{table}[ht]
\begin{tabular}{l|l|l|l|l}
 & \textbf{$E_\mathrm{B}$ [kcal/mol]} & \textbf{$\Delta_{\rm H}$ [cm$^{-1}$]} &
  \textbf{$S/\hbar$} &\textbf{$\Phi$} \\
  \hline
  \textbf{TL$_{\rm MEP}^{a}$} & 3.94 & 15.4 & 5.887 & 76.482\\
  \textbf{TL$_{\rm MEP}^{b}$} & 3.95 & 14.1 & 5.912 & 81.921\\
  \hline
  \textbf{TL$_{\rm IP}^{a}$} & 3.49 & 16.7 & 5.792 & 76.975\\
  \textbf{TL$_{\rm IP}^{b}$} & 3.50& 16.9 & 5.794 & 75.858\\
  \hline
  \textbf{TL$_{\rm MI}^{a}$} & 3.90 & 18.2 & 5.768 & 72.262\\
  \textbf{TL$_{\rm MI}^{b}$} & 3.90 & 16.9 & 5.774 & 77.374\\
  \hline
  \hline
  $\langle{\rm TL}_{\rm ext}\rangle$& 3.8945 & 25.3 & 5.743 & 53.241 
\end{tabular}
\caption{Energy barriers $E_\mathrm{B}$, tunneling splittings
  $\Delta_{\rm H}$ (at $T = 25$~K and $N = 4096$), action $S/\hbar$
  and fluctuation factor $\Phi$ (in a.u. of time $\hbar / E_h$) for HT
  determined from TL PESs using MEP points only (TL$_{\rm MEP}$),
  instanton points only (TL$_{\rm IP}$), and a combination of MEP and
  instanton points (TL$_{\rm MI}$). TL$_{\rm x}^{a}$ and TL$_{\rm
    x}^{b}$ correspond to two NNs that are trained on different splits
  of the data. Note that the energy barrier for TL$_{\rm IP}$ is
  inaccurate, as expected, because the IP misses the transition state
  of the PES. The {\it ab initio} barrier at the CCSD(T)/aug-cc-pVTZ
  level of theory is 3.8948~kcal/mol.}
\label{tab:tl_mep_results}
\end{table}

\noindent
First, it is noted that for TL$_{\rm MEP}$ and TL$_{\rm MI}$ the
barrier $E_\mathrm{B}$ for HT agrees well with the target value of
3.8948~kcal/mol (\textit{ab initio} CCSD(T)/aug-cc-pVTZ level value)
whereas this is not the case for TL$_{\rm IP}$, as expected, because
the instanton path does not pass through the transition state of the
MEP (see Table~\ref{tab:tl_mep_results}). Also, despite starting from
MP2 information only, TL to the HL model yields considerably improved
tunneling splittings $\Delta_{\rm H}$ for all three models, ranging
from 15 to 18 cm$^{-1}$, compared to those from the PhysNet MP2 PES
(96 cm$^{-1}$).\\

\noindent
Considering the action $S/\hbar$ from TL$_{\rm MEP}$, TL$_{\rm IP}$
and TL$_{\rm MI}$ it is seen that it progressively approaches that
from TL$_{\rm ext}$. For TL$_{\rm MEP}$ the action overshoots the
target value from TL$_{\rm ext}$ by $\sim 0.15$ which leads to an
error of $\sim 15$ \% in the splitting because $S/\hbar$ appears in
the exponential factor in Eq. \ref{eq:rpi_splitting}. Conversely, with
a difference of 0.04 compared with TL$_{\rm ext}$, the error for
TL$_{\rm MI}$ due to $S/\hbar$ is only $\sim 4$ \%. The influence of
$\Phi$ on the difference between TL$_{\rm ext}$ and the three models
considered in Table \ref{tab:tl_mep_results} is minor because for all
of them $\Phi$ is uniformly too large by $\sim 40$ \% compared with
that from TL$_{\rm ext}$.\\

\noindent
Table \ref{tab:tl_mep_results} suggests that a {\it
  combination} of information from the MEP and the IP used for transfer learning, i.e. TL$_{\rm MI}$, yields $E_{\rm B}$ and $\Delta_{\rm H}$ closest to the results from TL$_{\rm ext}$. However, the tunneling splittings still differ by more
than 10 \% from the target value obtained on the TL$_{\rm ext}$ PESs.
Considering the actions $S/\hbar$ and fluctuations $\Phi$ for all the
transfer learned PESs in Table~\ref{tab:tl_mep_results} it is found
that in particular the values of $\Phi$, which are sensitive to
fluctuations around the IP, differ considerably from that on TL$_{\rm
  ext}$. Hence as a last improvement points \emph{along} the MEP and IP are
combined with structures \emph{around} the IP.\\

\begin{table}[h!]
\begin{tabular}{c|c|c|ccc}
& $N_{\rm data}$ & \textbf{$E_{\rm B}$} & \textbf{$\Delta_{\rm H}$} &\textbf{$S/\hbar$} &\textbf{$\Phi$} \\\hline
MP2 & 70k & 2.7889 & 96.3 & 4.502 & 42.770\\\hline
 $\langle{\rm TL}_0\rangle$& 25&	$3.8925\pm0.0119$	&	$23.4\pm1.9$	&	$5.749\pm0.026$ & $57.405\pm4.486$	\\
 $\langle{\rm TL}_1\rangle$	& 50&	$3.8974\pm0.0161$	&	$24.9\pm1.1$	&	$5.764\pm0.01$2	&	$53.096\pm2.375$		\\
$\langle{\rm TL}_2\rangle$& 100&	$3.8941\pm0.0021$	&	$25.2\pm0.5$	&	$5.748\pm0.005$	&	$53.053\pm1.156$	\\
$\langle{\rm TL}_{\rm ext}\rangle$& 862	&	$3.8945\pm0.0006$	&	$25.3\pm0.2$	&	$5.743\pm0.002$	&	$53.241\pm0.454$\\
\hline
 $\langle{\rm TL}_{\rm EB1}\rangle$& 25 &$3.9025\pm0.0232$	&	$23.7\pm1.1$	&	$5.740\pm0.013$&	$57.099\pm3.249$	\\
 $\langle{\rm TL}_{\rm EB2}\rangle$& 25&	$3.9041\pm0.0189$	&	$23.5\pm2.1$	&	$5.733\pm0.025$&	$58.241\pm4.850$
\end{tabular}
\caption{Energy barriers $E_\mathrm{B}$ (kcal/mol), tunneling
  splittings $\Delta_{\rm H}$ (in cm$^{-1}$) at $T = 25$~K and with $N
  = 4096$, action $S/\hbar$ and fluctuation factor $\Phi$ (in a.u. of
  time $\hbar / E_h$) for malonaldehyde determined from TL$_{\rm
    EB}$. The {\it ab initio} barrier at the CCSD(T)/aug-cc-pVTZ level
  of theory is 3.8948~kcal/mol.}
\label{tab:evbased}
\end{table}

\begin{figure}[h!]
\centering
\includegraphics[width=0.8\textwidth]{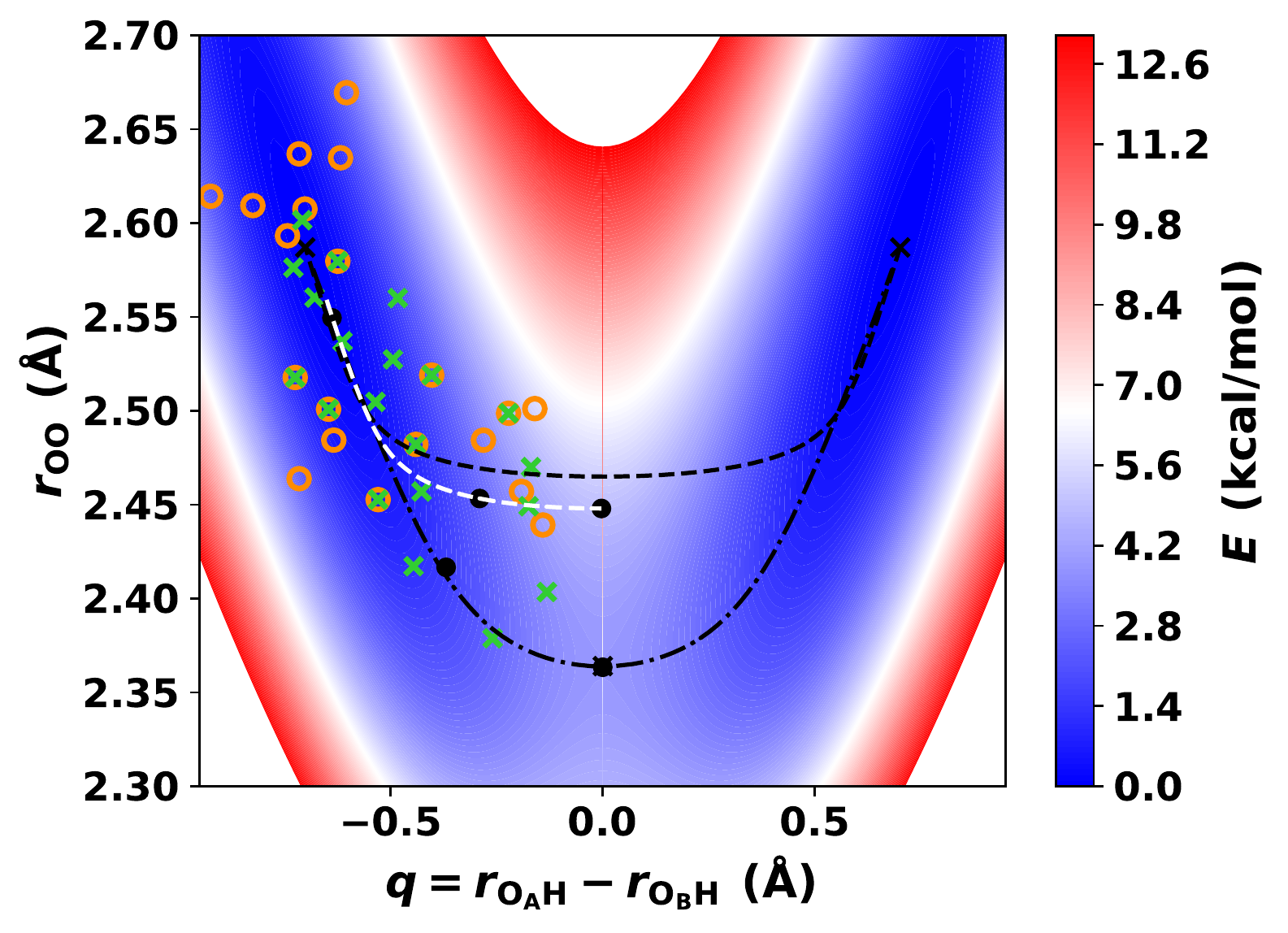}
\caption{The data sets used for TL$_{\rm EB1}$ (orange circles) and
  TL$_{\rm EB2}$ (green crosses) are shown on a 2D projection of the
  PES spanned by the O--O distance and the reaction coordinate $q =
  r_{\rm O_AH} - r_{\rm O_BH}$. Both data set contain 25
  geometries. The MEP and the IP (as determined on the TL$_{\rm ext}$
  PES) are marked with a dash-dotted and a dashed line,
  respectively. The IP on the MP2 PES is illustrated as white dashed
  line.}
\label{fig:evbased_tl1}
\end{figure}

\noindent
For two final, evidence-based TLs (TL$_{\rm EB1}$ and TL$_{\rm EB2}$),
a set of 25 data points was generated as follows. A total of 5 points
was selected along the MEP and IP (one close to the minimum and 2
points along the PhysNet MP2 PES MEP and IP each, see black points in
Figure \ref{fig:evbased_tl1}). These were supplemented by 20
geometries from NMS around the equilibrium structure and the IP that
are selected by means of an RMSD criterion.  For TL$_{\rm EB1}$
(orange circles in Figure~\ref{fig:evbased_tl1}), the geometries with
largest RMSD are selected (single geometries which occupy the same
$r_{\rm OO}$ and $q$ coordinates are eliminated). For TL$_{\rm EB2}$
(green crosses in Figure~\ref{fig:evbased_tl1}), besides the RMSD
criterion, the geometries were selected to cover the
\textit{important} configurational space more regularly (as judged by
Figure~\ref{fig:evbased_tl1}).\\

\noindent
For both sets of points the corresponding CCSD(T)/aug-cc-pVTZ
energies, forces and dipole moments were used for TL, resulting in two
sets of transfer learned PESs: TL$_{\rm EB1}$ and TL$_{\rm EB2}$. For
both of them the barrier height (3.90 kcal/mol) is within error bars
of TL$_{\rm ext}$ for which it was 3.89 kcal/mol. The action $S /
\hbar$ for both EB-models agree with TL$_{\rm ext}$ within error
bounds although the fluctuations around the mean is larger by almost
an order of magnitude, see Table \ref{tab:evbased}. For the
fluctuation $\Phi$ the differences compared with TL$_{\rm ext}$ are
$\sim 5$ \%, commensurate with TL$_0$ and evidently improved over
those using MEP, IP or MI, see Table \ref{tab:tl_mep_results} which do
not train on geometries around the path. The tunneling splittings are
$\Delta_{\rm H, EB1} = 23.7 \pm 1.1$ cm$^{-1}$ and $\Delta_{\rm H,
  EB2} = 23.5 \pm 2.1$ cm$^{-1}$, both of which are close to/within
error bounds of the reference value $(25.3 \pm 0.2)$, see
Table~\ref{tab:evbased}. These results are comparable and slightly
better to those on TL$_0$ which also was based on only 25 points for
TL. However, the training data for TL$_{\rm EB1,EB2}$ are selected
based {\it entirely} on the PhysNet MP2 PES whereas TL$_0$ made use of
HL information in that it employed geometries along the IP of a PES
with corrected barrier. Hence, from a computational perspective, the
EB models are considerably more cost-effective.\\

\noindent
Overall, it is found that TL to the HL with 25 additional points
yields a barrier height that agrees with the full HL treatment and the
tunneling splitting differs by only $\sim 1$ cm$^{-1}$. Any further
improvement requires additional points. Based on the results in
Table~\ref{tab:evbased} it is expected that using a TL model trained
on fewer than 50 judiciously selected HL data points yields results
within 1\% of the HL reference TL$_{\rm ext}$. This needs also be
contrasted with an expected accuracy of instanton calculations for
tunneling splittings of $\sim 20$ \%.\\

\section{Discussion and Conclusions}
The present work aimed at developing a computationally efficient and
accurate road-map for how to improve a given LL model - which was
assumed to be ``comprehensive'' (here $7 \times 10^4$ MP2 energies and
forces were used) for the observable of interest - to a HL model by
providing a small amount of additional information at the higher level
of theory considering a particular observable. Here, the observable
was the tunneling splitting for HT/DT in malonaldehyde for which the
LL model (MP2/aug-cc-pVTZ) yielded $\Delta= 96$ cm$^{-1}$, compared
with a literature value of $\Delta_{\rm H} \sim 25$ cm$^{-1}$ from a
PIP-represented PES of CCSD(T)/aug-cc-pVTZ reference energies using a
range of methods for computing $\Delta_{\rm H/D}$, see Table~S7. Most HL models generated were based on TL using 10s to
100s of HL points and yield $\Delta \sim 25$ cm$^{-1}$ which is a
substantial improvement over the LL model and consistent with
computations in the literature at the same level of theory but
employing computationally much more demanding approaches. The
remaining differences between computations and experiments are due to
a) shortcomings of the CCSD(T) level of theory compared with a full CI
treatment, b) the incompleteness of the basis set, and c) inherent
semiclassical approximations of instanton theory (e.g. neglect of
coupling to overall molecule rotation and anharmonicity perpendicular
to the instanton path).\\

\noindent
A typically used shortcut is to optimize the instanton using a LL
\textit{ab initio} method, e.g. DFT or MP2, and then compute the
CCSD(T) properties along the path to correct the action $S /
\hbar$. Such a hybrid approach was assessed using the PhysNet MP2 PES
to optimize the instanton and then calculate the action $S / \hbar$ on
the TL$_{\rm ext}$
PES.\cite{Milnikov2003,Meisner2018dual,chiral,nitrene} The results are
summarized in Table~\ref{tab:hybrid_action} and illustrate that
although the hybrid approach is, in this case, able to infer the
correct value for $S / \hbar$, the TL approach additionally improves
$\Phi$. Using Equation~\ref{eq:rpi_splitting} with the action as
determined by the hybrid approach ($S/\hbar = 5.7401$) and the
fluctuation factor determined on the PhysNet MP2 PES ($\Phi =
42.7705$) yields $\Delta_{\rm H} \sim 31.5$~cm$^{-1}$ which
overestimates the value of 25.1 cm$^{-1}$ from TL$_{\rm ext}$. The TL
approach is thus able to provide a more accurate prediction of
$\Delta$ for a similar computational cost.\\

\begin{table}[h!]
\begin{tabular}{c|ccc|ccc}
  & \multicolumn{3}{c}{H} & \multicolumn{3}{c}{D} \\
\textbf{PES}& \textbf{$\Delta_{\rm H}$} &  \textbf{$S/\hbar$} &\textbf{$\Phi$} & \textbf{$\Delta_{\rm D}$}& \textbf{$S/\hbar$} &\textbf{$\Phi$}\\\hline
MP2	&	96.3	&	4.502	&	42.771	&	18.8	&	5.705	&	73.964	\\
Hybrid	&	31.5	&	5.740	&	42.771	&	4.4	&	7.276	&	73.964	\\
 $\langle{\rm TL}_{\rm EB2}\rangle$& 23.5	&	5.733	&	58.241	&	3.5	&	7.266 & 94.777	\\
TL$_{\rm ext}$	&	25.1	&	5.744	&	53.586	&	3.6	&	7.274	&	89.957	\\
$\langle{\rm TL}_{\rm ext}\rangle$& 25.3	&	5.743	&	53.241	&	3.7	&	7.273	&	89.350	\\\hline
Lit.\cite{jahr2020instanton}	&	25	& 6.129 & 37.794&	3.3	& 7.790 & 61.392 

\end{tabular}
\caption{Actions $S/\hbar$ and fluctuation factor $\Phi$ as obtained from
  the MP2 PhysNet PES (MP2), on a representative TL$_{\rm ext}$ PES
  (TL$_{\rm ext}$) and a hybrid approach optimizing the IP on the
  PhysNet MP2 PES and using TL$_{\rm ext}$ to obtain CCSD(T)
  properties along resulting IP\cite{jahr2020instanton}}
\label{tab:hybrid_action}
\end{table}

\begin{table}[]
\begin{tabular}{rrrrrrrr}
\textbf{Mode} & \textbf{MP2} &  \textbf{$\langle{\rm TL}_0\rangle$} & \textbf{$\langle{\rm TL}_1\rangle$} &  \textbf{$\langle{\rm TL}_2\rangle$}& \textbf{$\langle{\rm TL}_{\rm EB1}\rangle$} & \textbf{$\langle{\rm TL}_{\rm ext}\rangle$}  & \textbf{CCSD(T)} \\\hline
1	&	277.49	&	266.68	&	265.48	&	265.28	&	266.93	&	265.17	&	264.71	\\
2	&	286.59	&	285.58	&	283.96	&	280.88	&	286.01	&	283.77	&	281.85	\\
3	&	394.33	&	389.41	&	387.06	&	389.09	&	392.05	&	389.70	&	389.12	\\
4	&	514.07	&	501.82	&	502.93	&	503.23	&	505.62	&	505.50	&	505.07	\\
5	&	789.38	&	772.93	&	774.86	&	775.82	&	773.14	&	775.33	&	775.06	\\
6	&	888.62	&	885.17	&	886.17	&	886.92	&	885.18	&	886.39	&	886.07	\\
7	&	937.61	&	906.84	&	907.54	&	908.64	&	908.10	&	908.03	&	907.18	\\
8	&	1012.29	&	988.09	&	989.05	&	992.64	&	991.79	&	990.78	&	989.73	\\
9	&	1023.82	&	1005.95	&	1004.01	&	1002.71	&	1006.17	&	1002.93	&	1002.60	\\
10	&	1048.75	&	1039.62	&	1039.69	&	1038.19	&	1039.50	&	1037.56	&	1037.85	\\
11	&	1109.69	&	1104.46	&	1104.81	&	1102.64	&	1102.08	&	1102.20	&	1101.03	\\
12	&	1288.28	&	1274.86	&	1271.94	&	1272.65	&	1272.27	&	1273.41	&	1272.73	\\
13	&	1403.09	&	1400.18	&	1399.97	&	1401.56	&	1402.15	&	1401.73	&	1400.73	\\
14	&	1407.97	&	1404.80	&	1408.40	&	1407.17	&	1406.59	&	1408.81	&	1406.94	\\
15	&	1482.06	&	1458.89	&	1463.84	&	1463.95	&	1462.21	&	1467.70	&	1469.33	\\
16	&	1641.52	&	1624.37	&	1627.95	&	1630.14	&	1631.91	&	1633.35	&	1632.52	\\
17	&	1692.91	&	1681.91	&	1687.06	&	1691.27	&	1686.72	&	1693.59	&	1693.63	\\
18	&	3039.02	&	3003.37	&	2999.68	&	2999.67	&	3005.82	&	3000.03	&	3001.26	\\
19	&	3107.12	&	3176.46	&	3178.76	&	3180.11	&	3176.23	&	3176.74	&	3176.35	\\
20	&	3217.85	&	3229.25	&	3228.54	&	3228.74	&	3232.47	&	3226.75	&	3227.12	\\
21	&	3267.30	&	3259.40	&	3260.39	&	3262.98	&	3254.33	&	3263.25	&	3260.27	\\\hline

\textbf{MAE}	&	\textbf{14.62}	&	\textbf{3.01}	&	\textbf{1.93}	&	\textbf{1.54}	& \textbf{2.61} &	\textbf{0.89}	&		
\end{tabular}
\caption{Averaged harmonic frequencies calculated from PhysNet
  potentials and using \textit{ab initio} techniques are given in
  cm$^{-1}$. As judged from the MAE($\omega$) the PhysNet model
  trained on the Extended Data Set containing 862 geometries is the
  most accurate, followed by the TL$_2$ (100 data points), TL$_1$ (50
  data points) TL$_{\rm EB1}$ (25 data points) and TL$_0$ (25 data
  points). The \textit{ab initio} harmonic frequencies obtained at the
  MP2 level are shown for comparison.}
\label{tab:harmfreq}
\end{table}

\noindent
For calculating the tunneling splittings based on the instanton
approach it was found that an evidence-based approach starting from
MEP and IP on the LL PES, augmented with geometries drawn from a pool
of structures selected such that their RMSD is maximal with respect to
an existing set of structures requires of the order of 50 points at
the HL for TL. Therefore, only {\it local} and not global knowledge of
the PES is required as would, e.g. be necessary for fully
quantum-mechanical methods such as wavepackets.  The bottleneck to a
``direct'' {\it ab initio-}based instanton approach is typically the
calculation of $N$ Hessians, as these are rather expensive to
compute.\cite{Milnikov2003,chiral} Earlier work on instanton rate
theory combined with machine-learning techniques for the H + CH$_4$
and H + C$_2$H$_6$ reactions required $\sim 50$ energies and forces
and 8 Hessians in the training set to converge the rate constant to
within $1$~\% of the \textit{ab initio} result at
200~K.\cite{laude2018ab} Using TL, calculating any high-level \emph{ab
  initio} Hessians at all has been avoided.  As is demonstrated here,
this can significantly lower the computational expense with no loss of
accuracy.\\

\noindent
With regards to the accuracy of the TL PESs it is of interest to
compare their performance on out-of-sample structures. For this a test
set was generated from MD simulations at 700~K on one of the TL$_{\rm
  ext}$ PESs from which 100 geometries were randomly extracted. In
addition, 10 equally spaced off-grid geometries along the IP on the
same PES were selected. The CCSD(T)/aug-cc-pVTZ energies of these 110
geometries cover a range from $\sim 5$ to $40$~kcal/mol above the
global minimum. The energies for these structures were computed based
on TL$_{\rm ext}$ (most rigorous TL using 862 HL structures) and
TL$_{\rm EB1}$ (following the recommended procedure; TL with 25 HL
energies and forces), respectively, and the
[MAE$_{100}$($E$),MAE$_{10}$($E$)] for the two out-of-sample sets are
[0.21,0.004] kcal/mol and [0.34,0.005] kcal/mol. Notably, the energies
of the geometries used for TL$_{\rm ext}$ and TL$_{\rm EB1}$ only
cover a range 20 kcal/mol above the global minimum whereas the
out-of-sample energies reach twice as high, up to 40 kcal/mol above
the minimum. Hence, the out-of-sample structures contain true
predictions on the HL-PES. As a comparison, for the PIP PES, which
used energies only and no forces, the reported fitting errors
(i.e. in-sample) are 32 cm$^{-1}$ (0.09 kcal/mol) for energies below
2000~cm$^{-1}$ (5.7 kcal/mol) above the global minimum and
211~cm$^{-1}$ (0.6~kcal/mol) for energies up to 20000~cm$^{-1}$ (51.2
kcal/mol).\cite{wang2008full}\\
  
\noindent
TL as used in the present work - namely as a local refinement of a
LL-PES - can also be regarded as a variant of the more global
``morphing'' approach for PESs.\cite{MM.morphing:1999} It is therefore
of interest to consider in what way observables other than the
tunneling splitting change upon TL from LL to HL. For this, harmonic
frequencies were determined for a number of transfer learned PESs. The
harmonic frequencies averaged over the 10 individually trained NNs for
different TLs are reported in Table~\ref{tab:harmfreq}, where they are
compared with frequencies determined from CCSD(T)/aug-cc-pVTZ
calculations at the corresponding equilibrium structure of
malonaldehyde. As judged from the MAE($\omega$) the PhysNet model for
TL$_{\rm ext}$ is most accurate (MAE $< 1$~cm$^{-1}$), followed by
TL$_2$ (MAE $< 2$~cm$^{-1}$), TL$_1$ (MAE $\sim 2$~cm$^{-1}$)and
TL$_0$ (MAE $\sim 3$~cm$^{-1}$), as expected, and show a considerable
improvement over the MP2 frequencies. For TL$_{\rm EB1}$ the MAE is
$<3$~cm$^{-1}$. Thus, TL to the HL model also improves the shape of
the PES in degrees orthogonal to the two reaction coordinates
considered for the tunneling splitting. \\

\begin{figure}[h!]
\centering
\includegraphics[width=1.05\textwidth]{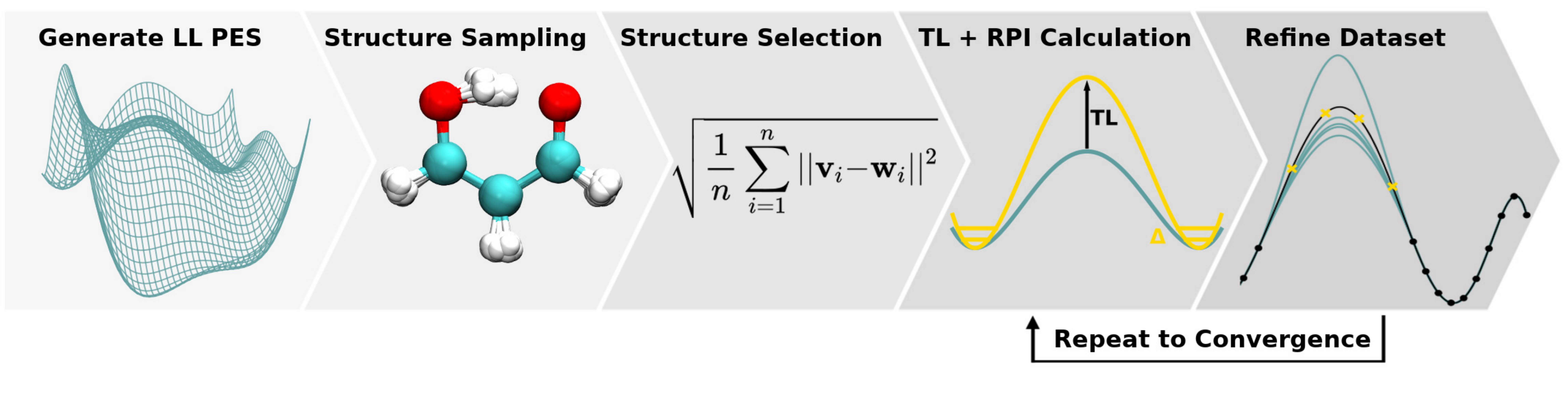}
\caption{Flowchart of the recommended TL + RPI procedure.}
\label{fig:flowchart}
\end{figure}

\noindent
In terms of a recommended procedure it is noted that the strategy
outlined in going from TL$_1$ to TL$_2$ (adaptive sampling/active
learning) can also be pursued recursively from a ``pool'' of
geometries generated from sampling the PhysNet MP2 PES. This procedure
can be repeated until convergence of the barrier height and the
tunneling splittings. The approach proposed for future application is
(see Figure~\ref{fig:flowchart}): i) create a LL-PES from a fine grid
(here $7 \times 10^4$ points) and train a ML model (here PhysNet) ii)
generate a pool of structures based on the LL-PES (including MEP,
Instanton, NMS) iii) choose $N \sim 25$ geometries following EB1/2 and
determine energies and forces from HL {\it ab initio} calculations iv)
perform TL and instanton calculations on the HL-PES v) refine the data
set using adaptive sampling and structures along the \textit{new}
IP. Then repeat TL and instanton calculations vi) repeat iv) and v)
until convergence. The present work demonstrates that following such a
road-map requires $\sim 50$ HL energy and force evaluations to
determine an accurate tunneling splitting of malonaldehyde which is
manifestly more efficient than previously explored approaches.\\

\noindent
In summary, given that LL models can be constructed efficiently even
for moderately sized molecules such as malonaldehyde or
larger,\cite{MM.rkhs:2020} the present work confirms that with
specific, evidence-based information grounded in physical
understanding of the process in question, several 10 points from a HL
treatment are sufficient to generate high-quality PESs for a target
observable which was the tunneling splitting in malonaldehyde in the
present work. The MAE($E$) for the TL-PESs trained on energies
spanning 20 kcal/mol above the global minimum is $\sim 0.3$ kcal/mol
on off-grid structures spanning 40 kcal/mol and the resulting harmonic
frequencies agree to within 1 to 3 cm$^{-1}$ with rigorous and very
time consuming normal mode analysis at the CCSD(T)/aug-cc-pVTZ level
of theory. The recommended approach deduced from the present work is
based on information about the MEP, the IP, and fluctuations around
the IP determined on the LL-PES for which HL calculations are required
for TL to determine the HL-PES. It is expected that - with
  suitable adaptations due to the particular observable considered -
  the present approach can also be applied to other observables that
  are computationally expensive to determine for a given PES, e.g. the
  quantum bound states of molecules or scattering cross sections for
  gas phase reactions from wavepacket propagation.\\

\section*{Acknowledgment}
This work has been financially supported by the Swiss National Science
Foundation NCCR-MUST (MM and JOR), the AFOSR, and the University of
Basel. \\

\bibliography{references,jeremy_refs}
\end{document}


\date{\today}

\begin{table}[ht]
\begin{tabular}{r|ccc}
\textbf{\textit{N} \textbackslash \textit{T}} & \textbf{50~K} & \textbf{25~K} & \textbf{12.5~K} \\\hline
512	&	96	&	94	&	87	\\
1024	&	97	&	96	&	94	\\
2048	&	97	&	96	&	96	\\
4096	&	97	&	96	&	96	
\end{tabular}
\caption{Tunneling splittings in cm$^{-1}$ for MA
  determined from the PhysNet MP2 PES\cite{mm.ht:2020}. Convergence is
  tested by looking down the diagonals of the table from which the
  optimal working effective temperature and $N$ can be
  determined.\cite{tunnel} Here 25~K and $N\ge1024$ is seen to be
  sufficient.}
\label{sitab:mp2_splittings}
\end{table}

\begin{table}[h!]
\begin{tabular}{c|c|c|ccc}
& $N_{\rm data}$ & \textbf{$E_{\rm B}$} & \textbf{$\Delta_{\rm D}$} &\textbf{$S/\hbar$} &\textbf{$\Phi$}\\\hline
MP2 & 70k & 2.7889 & 18.8& 5.705  & 73.964\\\hline
$\langle{\rm TL}_0\rangle$& 25&	$3.8925\pm0.0119$	&	$3.3\pm0.3$	&	$7.281\pm0.030$	&	$97.876\pm7.794$\\
$\langle{\rm TL}_1\rangle$	& 50&	$3.8974\pm0.0161$	&	$3.5\pm0.2$	&	$7.297\pm0.014$	&	$90.997\pm4.188$\\
$\langle{\rm TL}_2\rangle$& 100&	$3.8941\pm0.0021$	& $3.6\pm0.1$	&	$7.279\pm0.006$	&	$89.754\pm2.126$\\
$\langle{\rm TL}_{\rm ext}\rangle$& 862	&	$3.8945\pm0.0006$	&	$3.7\pm0.0$	&	$7.273\pm0.002$	&	$89.350\pm0.762$	
\end{tabular}
\caption{
  Energy barriers $E_\mathrm{B}$ (kcal/mol), tunneling splittings
  $\Delta$ (in cm$^{-1}$) at $T = 25$~K and with $N = 4096$, action
  $S/\hbar$ and fluctuation factor $\Phi$ for deuterated MA determined from
  TL$_{\rm ext}$ and TL$_{0,1,2}$ PESs. The {\it ab initio} barrier
  height for HT at the CCSD(T)/aug-cc-pVTZ level of theory is 3.8948~kcal/mol.}
\label{sitab:all_tl_splittings_D}
\end{table}

\begin{table}[h]
\begin{tabular}{c|r|rrr|rrr}
 \multicolumn{2}{c}{} & \multicolumn{3}{c}{H} & \multicolumn{3}{c}{D} \\
\textbf{TL$_{\rm ext}^\#$} & \textbf{$E_{\rm B}$} & \textbf{$\Delta_0$} &\textbf{$S/\hbar$} &\textbf{$\Phi$} & \textbf{$\Delta_0$} &\textbf{$S/\hbar$} &\textbf{$\Phi$}\\\hline
0	&	3.8938	&	25.1	&	5.744	&	53.586	&	3.64	&	7.273	&	89.957	\\
1	&	3.8942	&	25.4	&	5.742	&	52.925	&	3.70	&	7.272	&	88.696	\\
2	&	3.8942	&	25.6	&	5.744	&	52.513	&	3.71	&	7.274	&	88.308	\\
3	&	3.8938	&	24.9	&	5.740	&	54.138	&	3.61	&	7.270	&	91.017	\\
4	&	3.8939	&	25.3	&	5.740	&	53.238	&	3.69	&	7.270	&	89.014	\\
5	&	3.8953	&	25.1	&	5.745	&	53.418	&	3.65	&	7.275	&	89.498	\\
6	&	3.8943	&	25.4	&	5.743	&	52.959	&	3.69	&	7.273	&	88.870	\\
7	&	3.8953	&	25.4	&	5.743	&	52.943	&	3.67	&	7.273	&	89.321	\\
8	&	3.8953	&	25.1	&	5.744	&	53.542	&	3.65	&	7.275	&	89.723	\\
9	&	3.8950	&	25.4	&	5.741	&	53.142	&	3.69	&	7.271	&	89.097	\\\hline
$\langle{\rm TL}_{\rm ext}\rangle$	&	3.8945	&	25.3	&	5.743	&	53.241	&	3.67	&	7.273	&	89.350	\\
$\sigma$	&	0.0006	&	0.2	&	0.002	&	0.454	&	0.03	&	0.002	&	0.762
\end{tabular}
\caption{Energy barriers $E_B$ (kcal/mol), tunneling splittings
  $\Delta_0$ (at $T = 25$~K and $N = 4096$ given in cm$^{-1}$), action
  $S/\hbar$ and fluctuation factor $\Phi$ for MA and deuterated MA
  determined from TL$_{\rm ext}$ PES using an extended data set.}
\label{sitab:tl_extended}
\end{table}


\begin{table}[h]
\begin{tabular}{c|r|rrr|rrr}
 \multicolumn{2}{c}{} & \multicolumn{3}{c}{H} & \multicolumn{3}{c}{D}
 \\ \textbf{TL$_0^\#$} & \textbf{$E_{\rm B}$} & \textbf{$\Delta_0$}
 &\textbf{$S/\hbar$} &\textbf{$\Phi$} & \textbf{$\Delta_0$} &\textbf{$S/\hbar$}
 &\textbf{$\Phi$}\\\hline 0 & 3.8877 & 21.5 & 5.760 & 61.639 & 3.04 &
 7.292 & 105.980 \\ 1 & 3.8935 & 26.6 & 5.696 & 52.768 & 3.88 & 7.225
 & 88.422 \\ 2 & 3.8841 & 22.2 & 5.745 & 60.600 & 3.18 & 7.276 &
 102.895 \\ 3 & 3.8976 & 24.6 & 5.749 & 54.460 & 3.49 & 7.284 & 93.066
 \\ 4 & 3.8844 & 21.4 & 5.745 & 62.833 & 3.05 & 7.274 & 107.429 \\ 5 &
 3.8926 & 25.1 & 5.781 & 51.690 & 3.50 & 7.318 & 89.771 \\ 6 & 3.8904
 & 25.4 & 5.736 & 53.338 & 3.65 & 7.266 & 90.392 \\ 7 & 3.8771 & 22.2
 & 5.734 & 61.140 & 3.18 & 7.264 & 103.839 \\ 8 & 3.9214 & 23.8 &
 5.793 & 53.925 & 3.37 & 7.336 & 91.624 \\ 9 & 3.8958 & 21.8 & 5.745 &
 61.659 & 3.10 & 7.277 & 105.346 \\\hline $\langle{\rm TL}_0\rangle$ &
 3.8925 & 23.4 & 5.749 & 57.405 & 3.34 & 7.281 & 97.876\\ $\sigma$&
 0.0119 & 1.9 & 0.026 & 4.486 & 0.28 & 0.030 & 7.794
\end{tabular}
\caption{Energy barriers $E_B$ (kcal/mol), tunneling splittings
  $\Delta_0$ (at $T = 25$~K and $N = 4096$ given in cm$^{-1}$), action
  $S/\hbar$ and fluctuation factor $\Phi$ for MA and deuterated MA
  determined from TL$_0$ PES using a data set containing 50
  structures.}
\label{sitab:tl_small0}
\end{table}


\begin{table}[h]
\begin{tabular}{c|r|rrr|rrr}
 \multicolumn{2}{c}{} & \multicolumn{3}{c}{H} & \multicolumn{3}{c}{D} \\
\textbf{TL$_1^\#$} & \textbf{$E_{\rm B}$} & \textbf{$\Delta_0$} &\textbf{$S/\hbar$} &\textbf{$\Phi$} & \textbf{$\Delta_0$} &\textbf{$S/\hbar$} &\textbf{$\Phi$}\\\hline
0	&	3.8526	&	26.3	&	5.783	&	49.262	&	3.75	&	7.319	&	83.824	\\
1	&	3.9004	&	24.4	&	5.759	&	54.272	&	3.46	&	7.291	&	93.226	\\
2	&	3.9002	&	25.5	&	5.750	&	52.310	&	3.61	&	7.283	&	90.000	\\
3	&	3.9056	&	24.1	&	5.768	&	54.546	&	3.38	&	7.301	&	94.446	\\
4	&	3.9029	&	27.1	&	5.748	&	49.444	&	3.84	&	7.279	&	84.932	\\
5	&	3.9019	&	24.2	&	5.766	&	54.405	&	3.45	&	7.299	&	92.843	\\
6	&	3.9099	&	25.0	&	5.783	&	51.920	&	3.54	&	7.320	&	88.713	\\
7	&	3.8972	&	23.4	&	5.756	&	56.820	&	3.33	&	7.288	&	97.192	\\
8	&	3.9022	&	24.4	&	5.769	&	53.811	&	3.48	&	7.302	&	91.769	\\
9	&	3.9014	&	24.5	&	5.757	&	54.174	&	3.47	&	7.288	&	93.026	\\\hline
$\langle{\rm TL}_1\rangle$	&	3.8974	&	24.9	&	5.764	&	53.096	&	3.53	&	7.297	&	90.997	\\
$\sigma$	&	0.0161	&	1.1	&	0.012	&	2.375	&	0.16	&	0.014	&	4.188
\end{tabular}
\caption{Energy barriers $E_B$ (kcal/mol), tunneling splittings
  $\Delta_0$ (at $T = 25$~K and $N = 4096$ given in cm$^{-1}$), action
  $S/\hbar$ and fluctuation factor $\Phi$ for MA and deuterated MA
  determined from TL$_1$ PES using a data set containing 50
  structures.}
\label{sitab:tl_small1}
\end{table}


\begin{table}[h]
\begin{tabular}{c|r|rrr|rrr}
 \multicolumn{2}{c}{} & \multicolumn{3}{c}{H} & \multicolumn{3}{c}{D} \\
\textbf{TL$_2^\#$} & \textbf{$E_{\rm B}$} & \textbf{$\Delta_0$} &\textbf{$S/\hbar$} &\textbf{$\Phi$} & \textbf{$\Delta_0$} &\textbf{$S/\hbar$} &\textbf{$\Phi$}\\\hline
0	&	3.8950	&	25.1	&	5.754	&	53.109	&	3.60	&	7.285	&	90.133	\\
1	&	3.8974	&	24.8	&	5.753	&	53.760	&	3.55	&	7.285	&	91.390	\\
2	&	3.8931	&	24.8	&	5.742	&	54.284	&	3.62	&	7.272	&	90.596	\\
3	&	3.8920	&	25.9	&	5.751	&	51.488	&	3.72	&	7.282	&	87.312	\\
4	&	3.8922	&	26.0	&	5.744	&	51.648	&	3.74	&	7.273	&	87.720	\\
5	&	3.8932	&	25.5	&	5.746	&	52.648	&	3.68	&	7.276	&	88.701	\\
6	&	3.8942	&	24.8	&	5.753	&	53.636	&	3.55	&	7.285	&	91.271	\\
7	&	3.8968	&	25.5	&	5.752	&	52.316	&	3.67	&	7.283	&	88.488	\\
8	&	3.8959	&	25.6	&	5.743	&	52.528	&	3.73	&	7.272	&	87.869	\\
9	&	3.8917	&	24.4	&	5.745	&	55.116	&	3.48	&	7.274	&	94.056	\\\hline
$\langle{\rm TL}_2\rangle$	&	3.8941	&	25.2	&	5.748	&	53.053	&	3.63	&	7.279	&	89.754	\\
$\sigma$	&	0.0021	&	0.5	&	0.005	&	1.156	&	0.09	&	0.006	&	2.126
\end{tabular}
\caption{Energy barriers $E_B$ (kcal/mol), tunneling splittings
  $\Delta_0$ (at $T = 25$~K and $N = 4096$ given in cm$^{-1}$), action
  $S/\hbar$ and fluctuation factor $\Phi$ for MA and deuterated MA
  determined from TL$_2$ PES using a data set containing 100
  structures.}
\label{sitab:tl_small2}
\end{table}

\clearpage
\begin{figure}[h!]
\centering
\includegraphics[width=0.8\textwidth]{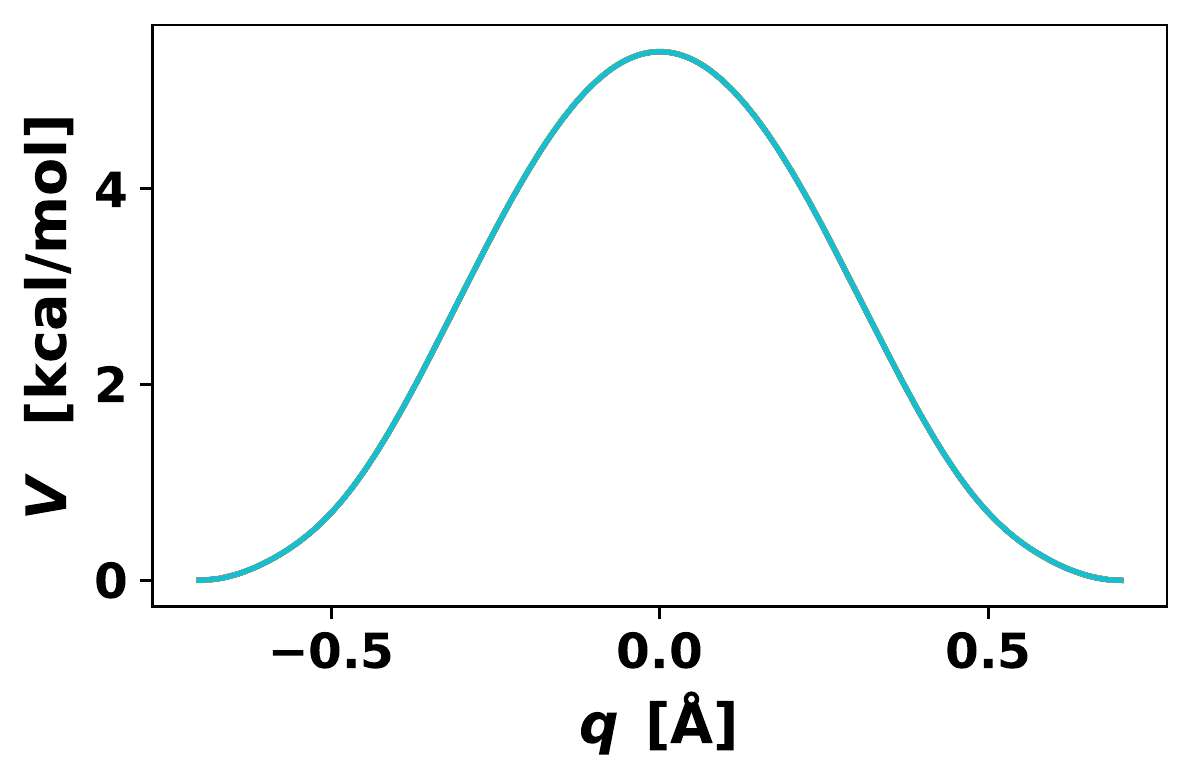}
\caption{Overlay of the instanton path obtained form the ten TLs using
  the extended data set, TL$_{\rm ext}$. $q$ corresponds to $q= r_{\rm
    O_AH} - r_{\rm O_BH}$.}
\label{sifig:instpath_ext}
\end{figure}

\begin{figure}[h!]
\centering
\includegraphics[width=0.8\textwidth]{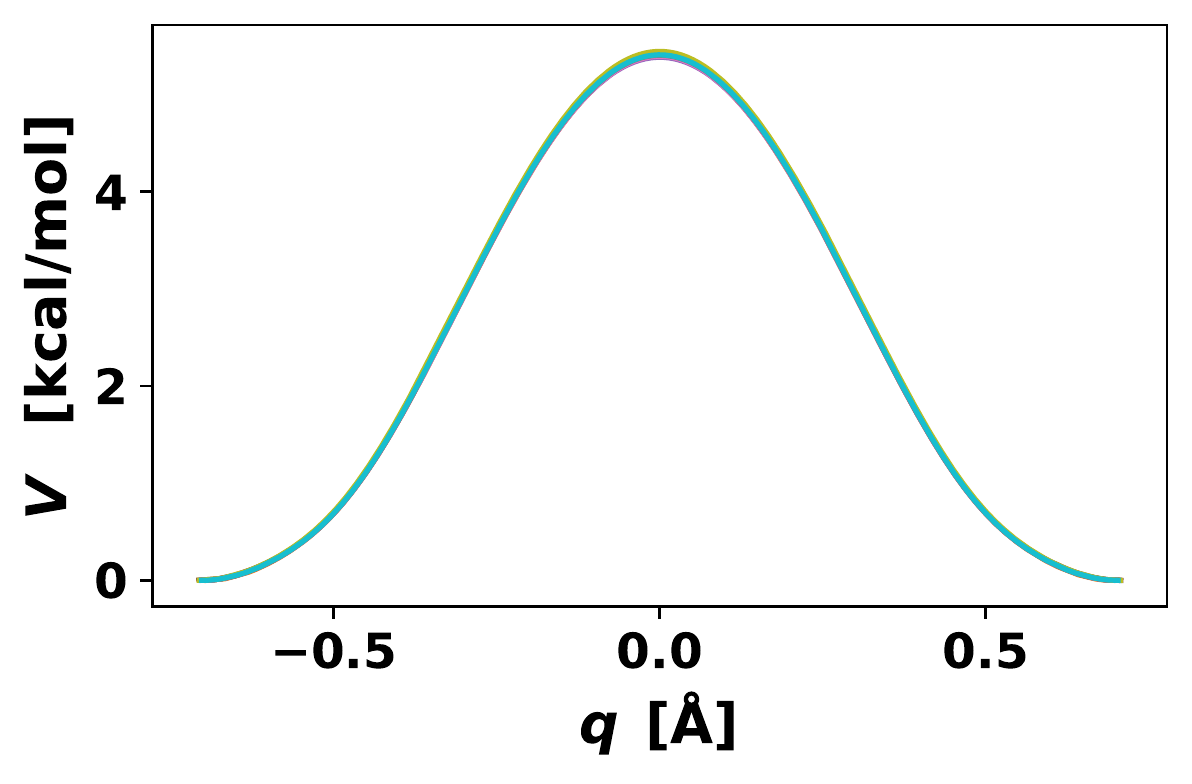}
\caption{Overlay of the instanton path obtained form the ten TLs from
  TL$_{\rm 0}$. $q$ corresponds to $q= r_{\rm O_AH} - r_{\rm O_BH}$.}
\label{sifig:instpath_tl0}
\end{figure}

\begin{figure}[h!]
\centering
\includegraphics[width=0.8\textwidth]{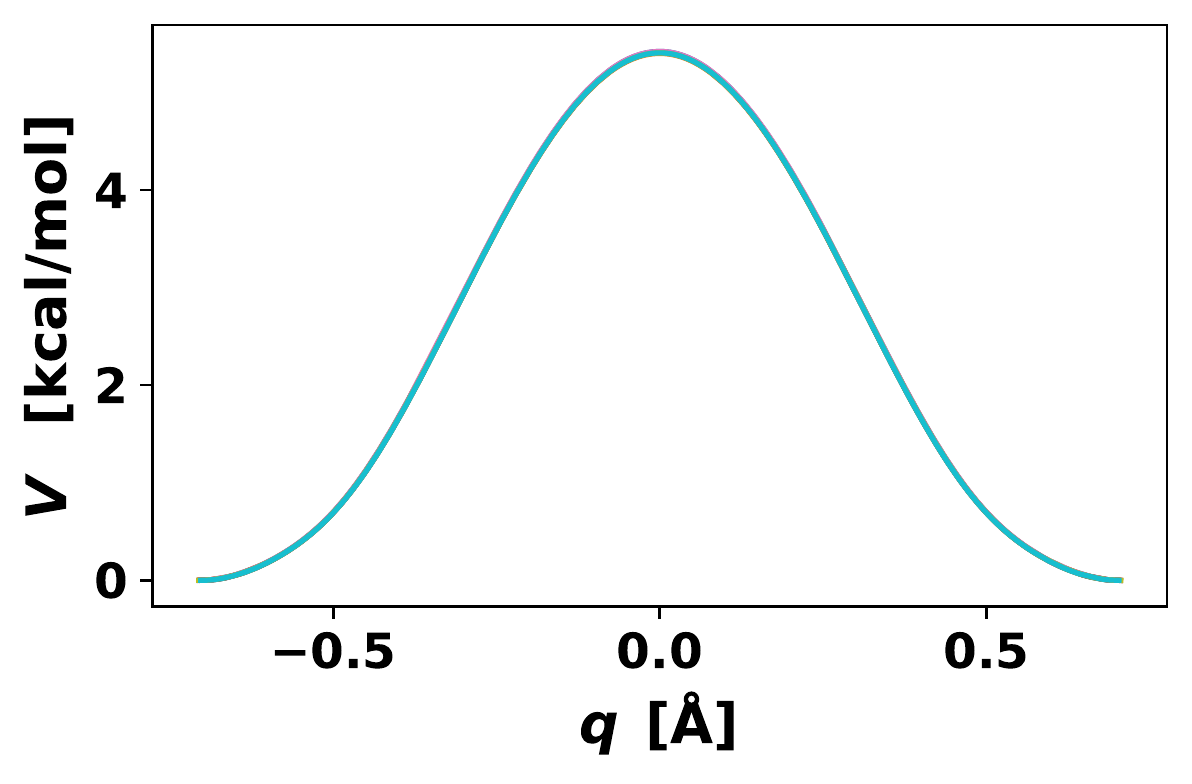}
\caption{Overlay of the instanton path obtained form the ten TLs from
  TL$_{\rm 1}$. $q$ corresponds to $q= r_{\rm O_AH} - r_{\rm O_BH}$.}
\label{sifig:instpath_tl1}
\end{figure}

\begin{figure}[h!]
\centering
\includegraphics[width=0.8\textwidth]{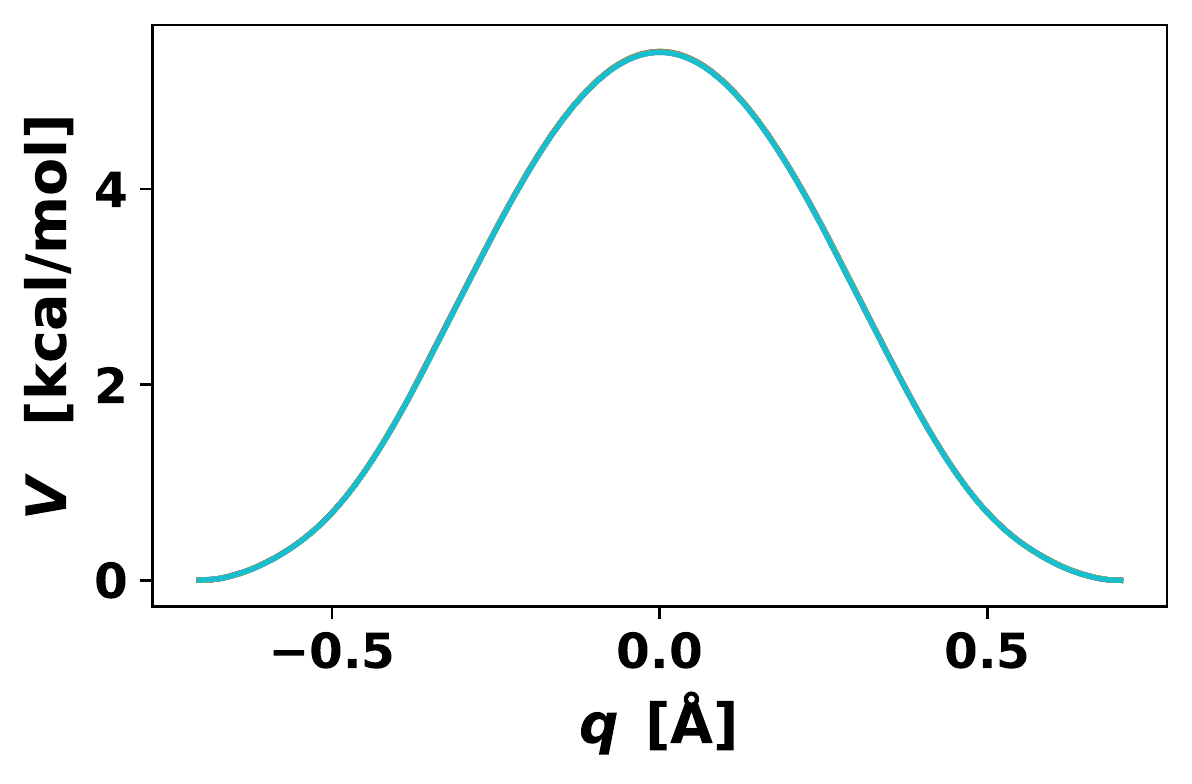}
\caption{Overlay of the instanton path obtained form the ten TLs from
  TL$_{\rm 2}$. $q$ corresponds to $q= r_{\rm O_AH} - r_{\rm O_BH}$.}
\label{sifig:instpath_tl2}
\end{figure}

\begin{figure}[h!]
\centering
\includegraphics[width=0.8\textwidth]{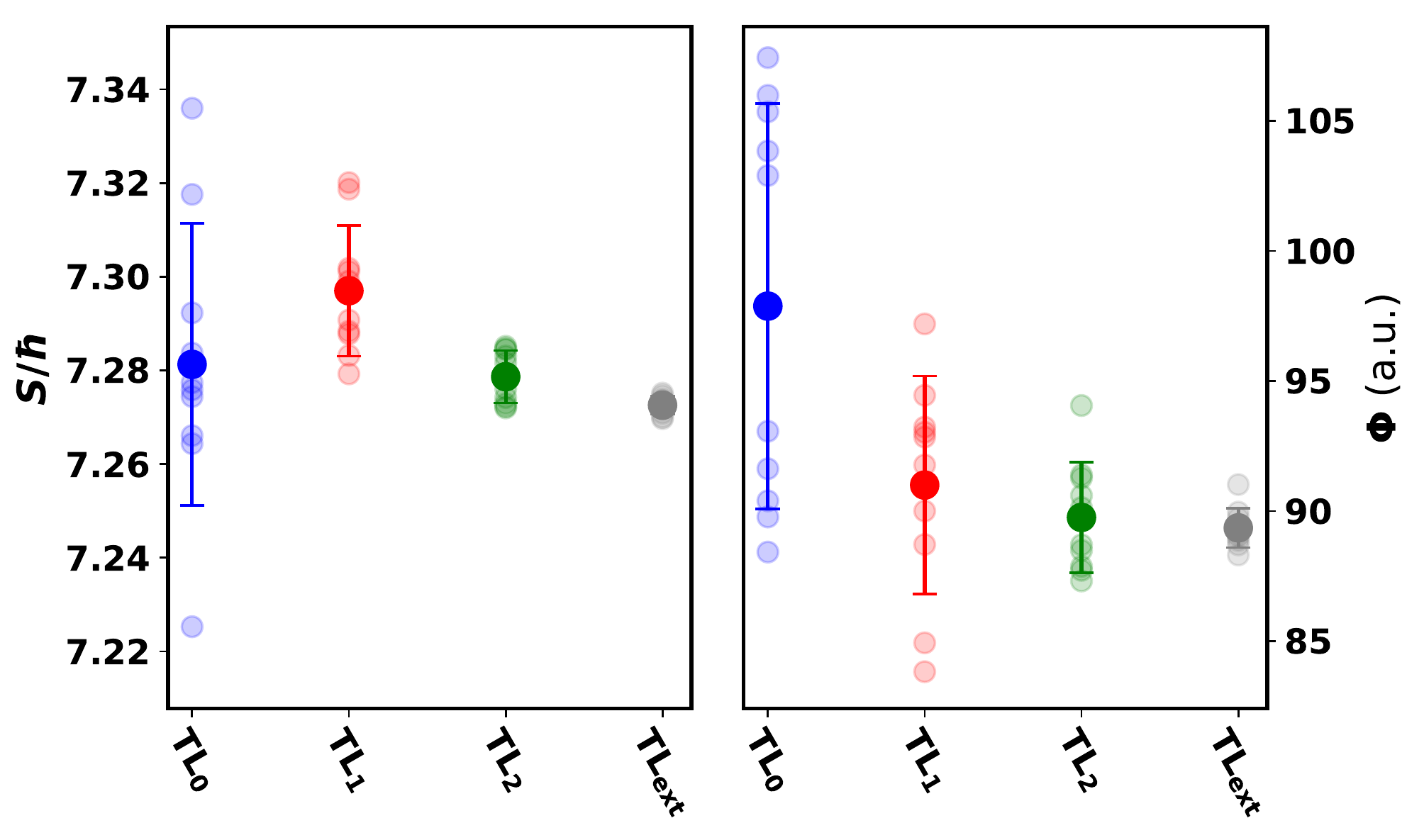}
\caption{Action $S/\hbar$ and fluctuation factor $\Phi$ from all TL
  PESs (transparent circles), the corresponding averages (opaque
  circle) and standard deviations (error bars as $\pm \sigma$) as
  obtained from TL$_0$ (blue), TL$_1$ (red), TL$_2$ (green) and
  TL$_{\rm ext}$ for DT.}
\label{sifig:fluctuation_factor}
\end{figure}

\begin{table}[h!]
\begin{threeparttable}
\scalebox{0.95}{
\begin{tabular}{l|clS[table-format=3.2]ccl}
 & \textbf{Ref.}& \textbf{PES} & \textbf{$E_{\rm B}$} & \textbf{Dynamics} & \textbf{H/D} & \textbf{$\Delta$ [cm$^{-1}$]} \\\hline
\textbf{Exp.} &\citenum{firth1991tunable,baba1999detection} & &  &  & H & 21.583 \\
&\citenum{baughcum1984microwave} & &  &  & D & 2.915 \\\hline
 \textbf{Theo.} & \citenum{sewell1995semiclassical}& FF\tnote{a} & 10.0 & semiclassical\tnote{b} & H & 21.8 \\
& \citenum{richardson2011ring} & FF\tnote{a,}\:\cite{sewell1995semiclassical}& 10.0 & RPI\tnote{i} & H & $51$\\
& \citenum{ben1999semiclassical}& AI\tnote{c} (B3LYP/double-$\zeta$) & 2.3 & semiclassical\tnote{b} & H & $21 \pm1$ \\
& \citenum{mil2004simple}& SI\tnote{d}\: (MP2/6-31G(d,p))\cite{yagi2001generation} & 3.6 & instanton & H & 30.7 \\
& \citenum{viel2007ground}& SI (MP2/6-31G(d,p))\cite{yagi2001generation} & 3.6 & POITSE\tnote{e} & H & $25.7 \pm 0.3$ \\
& \citenum{smedarchina2012rainbow} & AI (MC-QCISD/3)& 4.1 & rainbow instanton & H & $25.4$\\
& \citenum{mil2004simple}& AI\: (CCSD(T)/(a)VTZ) & 3.8 & instanton & H & 21 - 22 \\
& \citenum{wang2008full} & PIP\tnote{f}\: (CCSD(T)/aVTZ) & 4.1 & DMC & H & $21.6 \pm (2-3)$  \\
& \citenum{schroder2011theoretical} & PIP (CCSD(T)/aVTZ) & 4.1 & MCTDH\tnote{g} & H & $23.4$\\
& \citenum{hammer2011intramolecular} & PIP (CCSD(T)/aVTZ) & 4.1 & MCTDH & H & $23.8$\\
& \citenum{cvitas2016locating} & PIP (CCSD(T)/aVTZ) & 4.1 & RPI & H & \textcolor{red}{\bf 25}  \\
& \citenum{jahr2020instanton} & PIP (CCSD(T)/aVTZ) & 4.1 & RPI & H & \textcolor{red}{\bf 24.9}  \\
& \citenum{wu2016hydrogen} & PIP (CCSD(T)/aVTZ) & 4.1 & ti-QM\tnote{h} & H & $24.5$  \\
& \citenum{jahr2020instanton} & LASSO\tnote{j} (CCSD(T)(F12*)) & 4.0 & RPI & H & 19.3  \\\hline 
& \citenum{mil2004simple}& SI (MP2/6-31G(d,p))\cite{yagi2001generation} & 3.6 & instanton & D & 4.58 \\
& \citenum{viel2007ground}& SI (MP2/6-31G(d,p))\cite{yagi2001generation} & 3.6 & POITSE & D & $3.21 \pm 0.09$ \\
& \citenum{smedarchina2012rainbow} & AI (MC-QCISD/3)& 4.1 & rainbow instanton & D & $3.4$\\
& \citenum{mil2004simple}& AI (CCSD(T)/(a)VTZ) & 3.8 & instanton & D & 3.0 \\
& \citenum{wang2008full} & PIP (CCSD(T)/aVTZ) & 4.1 & DMC & D & $3.0\pm (2-3)$\\
& \citenum{cvitas2016locating} & PIP (CCSD(T)/aVTZ) & 4.1 & RPI & D & \textcolor{red}{\bf 3.4}  \\
 & & &  &  &  & 
\end{tabular}}
\caption{Summary of recent work on ground state tunneling splitting
  for MA.  $E_{\rm B}$ is given in kcal/mol. The results emphasized in
  red are supposed to be closest to the present results due to a
  similar PES and method to obtain the tunneling splittings (RPI).}
\begin{tablenotes}
    \item[a] experimental vibrational force field derived by Wilson et al.\cite{smith1983infrared,baughcum1981microwave}
    \item[b] semiclassical method based on the Makri-Miller model
    \item[c] \textit{Ab initio} 
    \item[d] Shepard interpolation
    \item[e] Monte Carlo projection operator, imaginary time spectral evolution
    \item[f] Permutationally invariant polynomial method
    \item[g] Multiconfiguration time-dependent Hartree approach
    \item[h] time-independent quantum mechanical method
    \item[i] Ring-polymer instanton method
    \item[j] least absolute shrinkage and selection operator
\end{tablenotes}
\label{tab:tab1}
\end{threeparttable}
\end{table}

\clearpage
\bibliography{references,jeremy_refs}